\PassOptionsToPackage{pdftex, xetex}{graphicx}
\PassOptionsToPackage{usenames,dvipsnames,svgnames,table}{xcolor}
\documentclass[letterpaper, 10pt, conference]{ieeeconf}
\IEEEoverridecommandlockouts
\makeatletter
\let\NAT@parse\undefined
\makeatother
\usepackage[sort,numbers,compress]{natbib}

\makeatother

\usepackage{amssymb, bm, bbm, mathtools,physics}
\usepackage[scr=boondoxo]{mathalfa}
\usepackage[usenames,dvipsnames,svgnames,table]{xcolor}
\usepackage[pdftex, xetex]{graphicx}
\usepackage[font=small]{caption}
\usepackage{setspace}
\usepackage{float, colortbl, tabularx, longtable, multirow, subcaption, environ, wrapfig, textcomp, nicefrac}
\usepackage{pgf, tikz, framed}
\usepackage[normalem]{ulem}
\usetikzlibrary{arrows,positioning,automata,shadows,fit,shapes,cd}
\usepackage[USenglish]{babel}
\usepackage{blindtext}
\usepackage{booktabs,colortbl}
\usepackage{pdfpages,pgffor}
\usepackage{lineno}
\usepackage{xargs}

\usepackage[final]{microtype}

\usepackage[T1]{fontenc}
\usepackage{newtxtext}

\usepackage{pythonhighlight}
\usepackage{fontawesome}
\usepackage{bibentry}
\usepackage{listings}
\definecolor{myred}{RGB}{153,51,51}
\definecolor{mygray}{RGB}{102,102,102}
\definecolor{darkgreen}{RGB}{60,160,60}
\definecolor{lightblue}{RGB}{60,60,200}
\definecolor{darkred}{RGB}{70,10,10}
\definecolor{darkblue}{RGB}{1,1,255}
\definecolor{yellow}{RGB}{255,255,0}
\definecolor{purple}{RGB}{127,0,255}
\definecolor{dark}{RGB}{1,1,1}
\definecolor{lightgray}{RGB}{230,230,230}
\definecolor{ucla_gold}{RGB}{255,232,0}
\definecolor{ucla_blue}{RGB}{50,132,191}
\definecolor{lightyellow}{rgb}{1.0, 1.0, 0.88}

\colorlet{mylinkcolor}{violet}
\colorlet{mycitecolor}{YellowOrange}
\colorlet{myurlcolor}{Aquamarine}

\definecolor{color_skyblue}{rgb}{0.01,0.39,0.75}

\usepackage[colorlinks=true]{hyperref}
\hypersetup{
    linkcolor  = RoyalBlue,
    citecolor  = RoyalBlue,
    urlcolor   = RoyalBlue,
    colorlinks = true,
    linktoc=all,
    breaklinks=true,
}
\usepackage{url}

\makeatletter
\def\th@plain{%
  \thm@headfont{\bfseries}
  \thm@notefont{}
  \normalfont 
}
\def\th@definition{%
  \thm@notefont{}
  \normalfont 
}
\makeatother

\makeatletter
\newcommand\numberprefix{}
\def\numberline#1{\hb@xt@\@tempdima{\numberprefix #1\hfil}}
\def\l@figure#1#2{\@tocline{0}{3pt plus2pt}{0pt}{5pc}{}%
  {\renewcommand\numberprefix{}#1~}{#2}}
\def\l@table#1#2{\@tocline{0}{3pt plus2pt}{0pt}{5pc}{}%
  {\renewcommand\numberprefix{}#1~}{#2}}
\makeatother

\usepackage[capitalize,nameinlink,noabbrev]{cleveref}
\crefformat{equation}{(#2#1#3)}
\crefmultiformat{equation}{(#2#1#3)}%
{ and~(#2#1#3)}{, (#2#1#3)}{ and~(#2#1#3)}
\crefformat{theorem}{Theorem~#2#1#3}
\crefformat{lemma}{Lemma~#2#1#3}
\crefformat{remark}{Remark~#2#1#3}
\crefformat{section}{Section~#2#1#3}
\crefformat{assumption}{Assumption~#2#1#3}
\crefformat{example}{Example~#2#1#3}
\crefformat{figure}{Figure~#2#1#3}
\crefformat{algorithm}{Algorithm~#2#1#3}
\crefrangeformat{section}{Sections~#3#1#4--#5#2#6}
\crefformat{assumption}{Assumption~#2#1#3}

\crefrangeformat{equation}{~(#3#1#4--#5#2#6)}
\crefrangeformat{figure}{Figures~#3#1#4--#5#2#6}
\crefrangeformat{assumption}{Assumptions~(#3#1#4--#5#2#6)}

\newcommand{\aed}[1]{\begin{aligned} #1 \end{aligned}}
\newcommand{\beq}[1]{\begin{equation}#1\end{equation}}

\newcommand{\rbr}[1]{\left(#1\right)}

\providecommand\f[2]{\ensuremath \frac{#1}{#2}}

\let\max\relax

\DeclareMathOperator*{\max}{\text{max}}

\DeclareMathOperator*{\argmax}{\text{argmax}}

\def \s {\sigma}

\def \l {\lambda}

\def \reals {\mathbb{R}}

\def \H {\text{H}}
\def \I {\text{I}}
\def \KL {\text{KL}}
\graphicspath{{./figures/}}

\usepackage[color=gray!30,textsize=tiny]{todonotes}

\usepackage{fancyhdr}
\usepackage{algpseudocode}
\usepackage{listings}
\usepackage{ifthen}

\makeatletter
\DeclareMathSizes{\@xpt}{9}{7}{5}
\makeatother

\title{\bf Active Scout: Multi-Target Tracking Using Neural Radiance Fields in Dense Urban Environments}

\author{Christopher D. Hsu$^{1,2}$ and Pratik Chaudhari$^{2}$
\thanks{
$^{1}$DEVCOM Army Research Laboratory
\href{mailto:christopher.d.hsu.civ@army.mil}{christopher.d.hsu.civ@army.mil}
}
\thanks{
$^{2}$Department of Electrical \&  Systems Engineering and
General Robotics, Automation, Sensing and Perception (GRASP) Laboratory at the University of Pennsylvania.
\href{mailto:chsu8@seas.upenn.edu}{chsu8@seas.upenn.edu},
\href{mailto:pratikac@seas.upenn.edu}{pratikac@seas.upenn.edu}}
}

\begin{document}

\maketitle


\thispagestyle{fancy}
\fancyhead{} 
\fancyfoot{} 
\fancyhead[C]{\ifthenelse{\value{page}=1}{Distribution Statement A. Approved for public release: distribution is unlimited.}{UNCLASSIFIED}}
\fancyfoot[C]{\ifthenelse{\value{page}=1}{}{UNCLASSIFIED}}
\pagestyle{fancy}
\setlength{\headheight}{12pt}
\setlength{\footskip}{12pt}

\begin{abstract}
We study pursuit-evasion games in highly occluded urban environments, e.g. tall buildings in a city, where a scout (quadrotor) tracks multiple dynamic targets on the ground. We show that we can build a neural radiance field (NeRF) representation of the city---online---using RGB and depth images from different vantage points. This representation is used to calculate the information gain to both explore unknown parts of the city and track the targets---thereby giving a completely first-principles approach to actively tracking dynamic targets. We demonstrate, using a custom-built simulator using Open Street Maps data of Philadelphia and New York City, that we can explore and locate 20 stationary targets within 300 steps. This is slower than a greedy baseline, which does not use active perception. But for dynamic targets that actively hide behind occlusions, we show that our approach maintains, at worst, a tracking error of 200m; the greedy baseline can have a tracking error as large as 600m. We observe a number of interesting properties in the scout's policies, e.g., it switches its attention to track a different target periodically, as the quality of the NeRF representation improves over time, the scout also becomes better in terms of target tracking. Code is available at \url{https://github.com/grasp-lyrl/ActiveScout}.



%

\end{abstract}


\section{Introduction}
\label{s:intro}

\begin{figure}[H]
    \centering
    \includegraphics[width=0.88\linewidth]{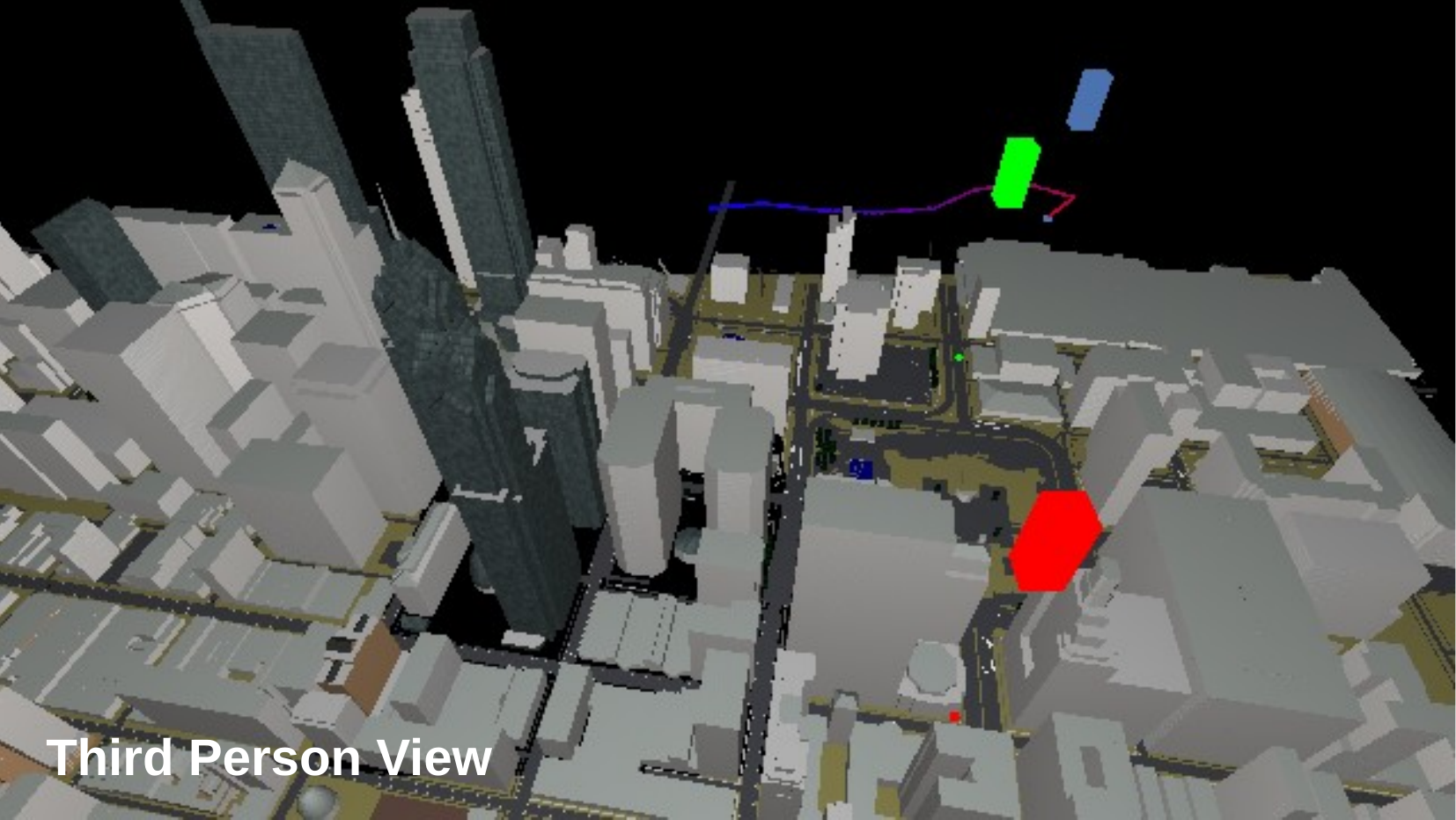}\\
    \vspace*{0.2em}
    \includegraphics[width=0.43\linewidth]{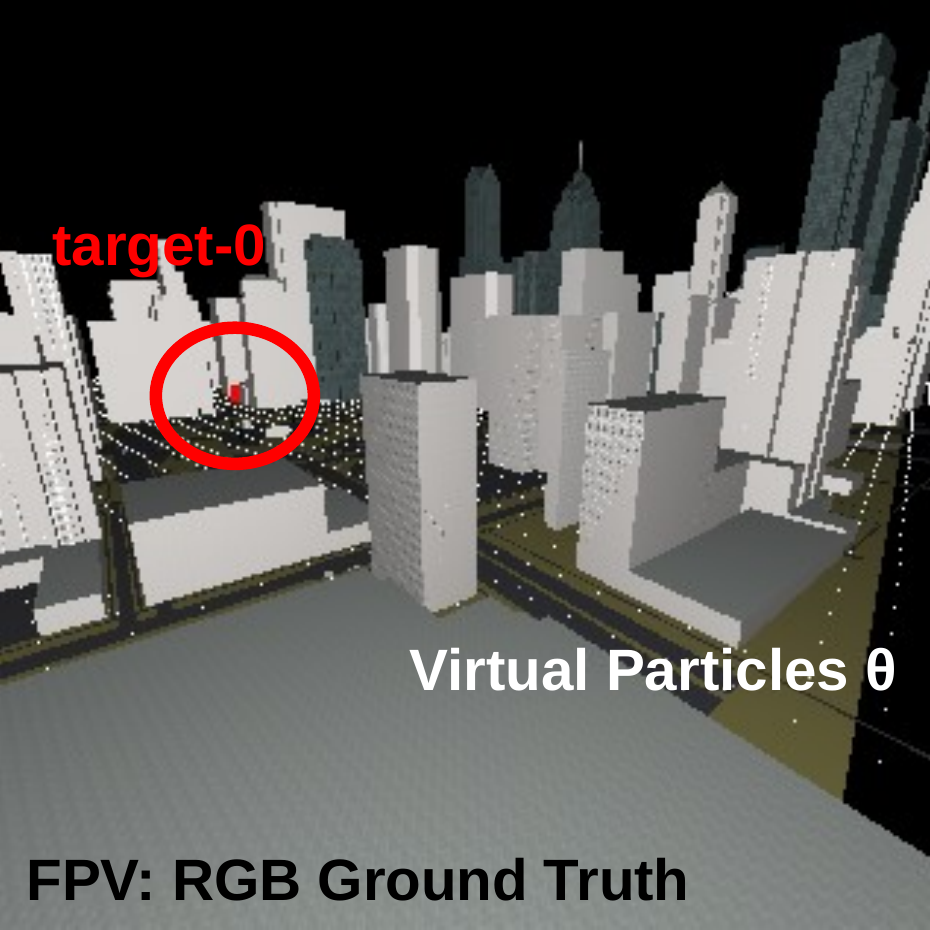}
    \includegraphics[width=0.43\linewidth]{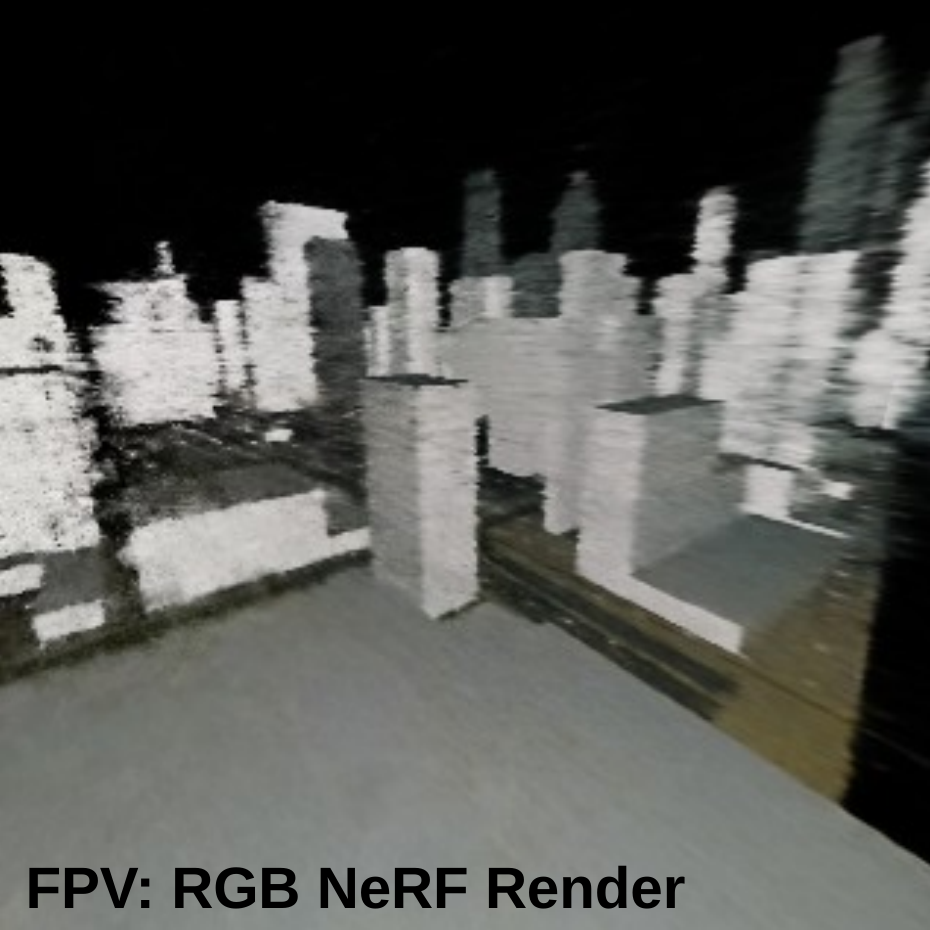}
    \includegraphics[width=0.88\linewidth]{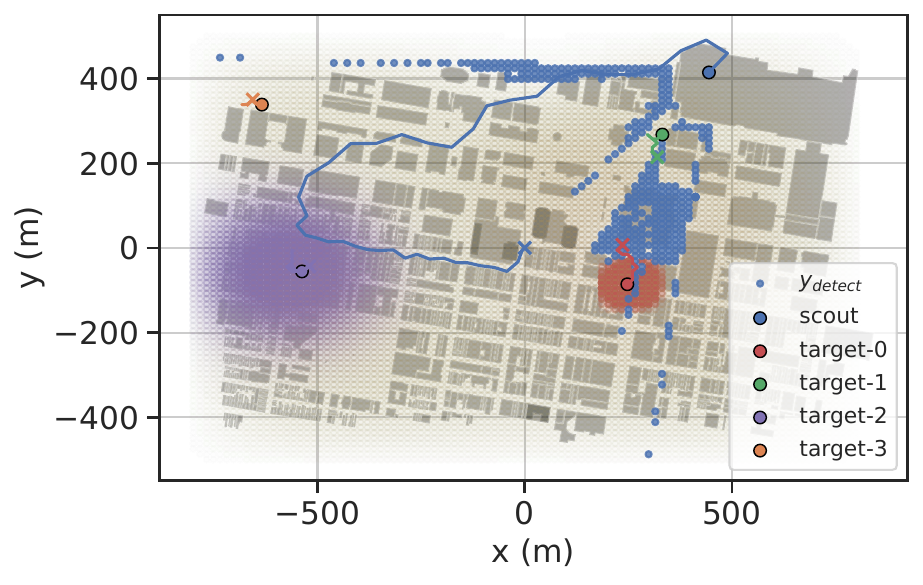}
    \caption{We present a snapshot of the scout (blue) in a Philadelphia scene with 4 targets. \textbf{Top:} third person view of the scene with hovering agent beacons. The red-blue line denotes the scout's trajectory history. \textbf{Middle:} On the left is the scout's first person view (minus the labels) where it can see a red `target-0' and virtually projected ground particles $\theta$ (white dots) that help update the target filter. On the right we show the synthesized NeRF rendering of that view after training. \textbf{Bottom:} 2D map showing the building footprints (grey), the scout and its current observation of the virtual particles $y_{\text{detect}}$ (blue scatter), and the targets. Each target has an associated filter (opacity denotes weight). In this snapshot, the red target's filter is updated due to being observed. The purple target was viewed in the past and its uncertainty spreads over time due to the motion model. The orange and green targets have not been viewed so their prior is still rather uniform over the scene (orange+green=brown).
    }
    \label{fig:1}
\end{figure}

Consider a game of cops and robbers in which a quadrotor scout (cop) must search for and track robbers within a city, see~\cref{fig:1}. Robbers actively avoid the scout by hiding in blind spots, or unknown parts of the environment. In this paper, we ask: what is the next best view for the scout to maximize its information of the targets' locations? We focus on three specific aspects: (1) If the scout does not have a map of the scene, how does it explore to build one on the fly? (2) How should the scout trade off between learning about the environment and tracking targets? (3) How should targets use blind spots created by occlusions to hide from the scout?

The contributions of this work are as follows. (a) Neural radiance fields (NeRFs) can be trained online to represent the history of color and depth images observed; they should be used to synthesize future views given a sample of future poses. As NeRFs are not probabilistic models, we show that we can build an ensemble from bootstrapped data and calculate a variance over color and depth. (b) Bayes filters can represent a history of detected target locations and we show how to incorporate them into the NeRF representation. 
(c) The computation of mutual information provides a seamless way of integrating exploration and tracking objectives. With a ranking of sample poses, we can select a pose and perform a dynamically feasible quadrotor trajectory to the selected waypoint. Finally, (d) we provide a policy for an active target that utilizes knowledge of scout's history of observations and moves to locations that appear occluded to the scout.

\section{Related Work}

Inspired by autonomous information gathering problems~\cite{wald1945}, the prosperous line of work called active perception~\cite{bajcsy2016revisiting} developed the central tenet that an active perceiver should take control actions that lead to informative observations and use this data in a tight feedback loop to select the next set of controls.
The problem discussed here embodies active perception and is rooted in adaptive sampling for which the goal is to choose the best option from a set of samples that minimize prediction uncertainty or maximization of some information gain~\cite{Hoffman2010}, in comparison to offline non-adaptive where the environment is static and the plan is computed offline~\cite{singh2009efficient}. Whether adaptive or non-adaptive, mutual information satisfies submodularity~\cite{Krause2011submodularity} which shows that selecting sequential sensors locations in a greedy fashion has a sensing quality that is provably close to the optimal sensing quality. This property holds for static scenes and has a flavor of the multi-agent sensor problem~\cite{HollingerSukhatme2014sampling}. It is unclear how this property holds for temporally dynamic scenes or processes of interest. Even so, other works have found success employing a greedy (myopic control) approach to dynamic multi-target problems~\cite{SpletzerTaylor2003,DamesKumar2017}. On the other hand, non-myopic control solutions have had success in dynamic information gathering problems~\cite{schlotfeldt2018anytime}.

In the aforementioned works, the environment for which the robots are deployed in are simplistic with simple measurements models, i.e. bearing or range measurements with noise. In the case of urban environments or occlusions, not much has been done, but works generally consider occlusions within the control optimization scheme as obstacles~\cite{Hausman2016Occlusion, Vanegas2018occlude} and these obstacles are often not intrusive to the tracking of the targets. There is work that considers urban environments~\cite{berry2019obstructed} and they incorporate the occlusions in their particle filter update. However, their sensor is stationary and uses simplistic measurement models.
We hypothesize that for an active perceiver in a complex environment with large occlusions (in our case due to tall buildings in a cityscape) more complex measurement and map representations are needed to test the efficacy of these works. In our past work~\cite{he2023active}, we argued that a neural radiance field (NeRF)~\cite{mildenhall2020nerf} is well-suited for active perception tasks for its ability to summarize multi-modal information, e.g. photometric and geometric, in a consistent fashion and synthesize new views to be able to calculate information-based objectives such as predictive information~\cite{bialek2001predictability}. Uncertainty quantification for next best view selection with radiance fields has also seen success on constrained scenes~\cite{jiang2023fisherrf}. On larger scenes NeRFs have been productive with the caveat of well constructed adjustments such as training on progressively different scales of data while also expanding the NeRF concurrently~\cite{xiangli2023bungeenerf}, or decomposing the scene into multiple NeRFs~\cite{tancik2022blocknerf}. In this work we use a fixed neural network size and online data which trades off high quality resolution reconstruction for a smaller memory footprint and speed of training. In fact, the focus of this work is the use of NeRFs as a mode to balance exploration and exploitation in target tracking in an unknown complex environment. 


However, NeRFs are only one part of the equation. Here we seek to solve the multi-target tracking problem. Past works have used all types of filters to localize targets but predominantly, kalman filters and their variants~\cite{Charrow2015thesis}, particle filters~\cite{SpletzerTaylor2003,Hoffman2010}, and probability hypothesis density (PHD) filters~\cite{DamesKumar2017, banerjee2024decentralized}. Single agent tracking with these methods is straightforward but the difficulty lies with multi-target tracking. Generally one would make multiple copies of the filter of choice to track each individual target. However, in the works that use the PHD filter, they forgo the assumption of data association, i.e. given a target the user can update the correct associated filter. In past works, this is a valid thought process to consider as range and bearing measurements of the targets lose vital information. In the context of this work, we employ a perception system, e.g. an rgbd (color and depth) camera to perform measurements. We assume that with our perceptual system we can distinguish between multiple targets and update the associated filter~\cite{Tokekar2014visualtracking}. 

Finally we want to mention another popular line of work in multi-target tracking: when you have a team of agents. Past works have employed graph neural networks~\cite{tzes2022graph}, joint or decentralized estimation over the information filter~\cite{Hoffman2010, schlotfeldt2018anytime, banerjee2024decentralized}, or multi-agent reinforcement learning algorithms~\cite{hsu2021scalable}. In this work, multi-agent teaming is not the focus. We recount the submodularity property of mutual information which states that the greedy selection of locations of subsequently added sensors is near optimal.

\section{Problem Formulation}
\label{s:problem}

\begin{table}[!htbp]
\centering
\footnotesize
\caption{Key quantities in the text.}
\renewcommand{\arraystretch}{1.25}
\begin{tabular}{cc}
\toprule
$\xi$ & scene\\
$\theta_t^{(i)}$ & location of the $i^{\text{th}}$ target at time $t$\\
$x_t \in \text{SE(3)}$ & location of the scout at time $t$\\
$x_{\text{past}} \equiv x_{0:t}$ & past locations of the scout\\
$x_{\text{future}} \equiv x_{t+\Delta t}$ & future location of the scout\\
$y^{(i)} = (y_{\text{rgb}}, y_{\text{depth}}, y_{\text{detect}}^{(i)})$ & RGB, depth images of the scene and \\
& detections of the $i^{\text{th}}$ target\\
$y_{\text{future}} \equiv y_{t+\Delta t}^{(i)}$ & future observation\\
$\H(\cdot), \I(\cdot)$ & Shannon entropy and mutual information\\
\bottomrule
\end{tabular}
\label{table}
\end{table}
\cref{table} is a summary of the the key quantities that will be introduced in the text that follows.
Denote the location of the scout (a quadrotor) by $x_t \in$ the special Euclidean group SE(3) and its dynamics by $\dot x = f(x_t, u_t)$ where $u_t$ denotes the control input; we will elaborate upon the dynamics model later. Locations of the $m$ mobile ground targets are $\theta^{(i)} \in \reals^2$. We will assume that the scout can localize itself, e.g. with GPS, i.e., $x_t$ is known perfectly. The scout receives RGB and depth images from the scene $\xi$ and, when the targets are not occluded, it can detect them in the images. Let $y^{(i)}_t = (y_{\text{rgb}}, y_{\text{depth}}, y_{\text{detect}}^{(i)})$ denote the observation received at time $t$ corresponding to the $i^{\text{th}}$ target; it consists of RGB and depth images from the scout's location and detections of the target in these images.

The scout searches and tracks the targets. We developed an active perception objective for such problems in~\cite{he2023active} and argued that an agent performing active perception should maximize the mutual information that past observations contain about future ones. Future observations are, of course, unavailable, and therefore such an agent should have the ability to synthesize new observations, i.e., a generative model. This representation would ideally be constructed incrementally using past observations. We set
\beq{
    u_{\text{future}} \in \argmax_{\varphi} \I(y_{\text{future}}; y_{\text{past}} \mid x_{\text{future}}).
    \label{eq:MIobj}
}
where $\varphi \equiv p(x_{\text{future}} \mid y_{\text{past}})$ is the probability distribution of the next scout location over which we are optimizing, and it depends upon the control $u_{\text{future}}$. Mutual information between two random variables $\I(y; x)$ is defined as
\[
    \aed{
        \I(y; x) &= \int \dd{y}\dd{x} p(x, y) \log \f{p(x, y)}{p(x) p(y)}\\
        &=\H(y) - \H(y \mid x),
    }
\]
where $\H$ is the Shannon entropy. It is equal to the Kullback-Liebler (KL) divergence $\KL(p(y \mid x), p(y))$ averaged over all possible realizations of $x$. In our case, the mutual information characterizes the discrepancy between the scout's future observations given its future locations and the scout's observations given its past locations/observations. The scout takes control actions that maximize this discrepancy, i.e., it maximizes the information gain to take control actions that provide new information about the scene and the targets. To calculate the mutual information, the scout must also be able to sample future observations $y_{\text{future}}$ given past ones (this means, both how images from the scene $\xi$ will look and where the targets might be detected in these images). We will discuss how to do this in the following sections.

\section{Methodology}
\label{s:method}

\begin{figure}[!htpb]
    \centering
    \includegraphics[width=0.75\linewidth]{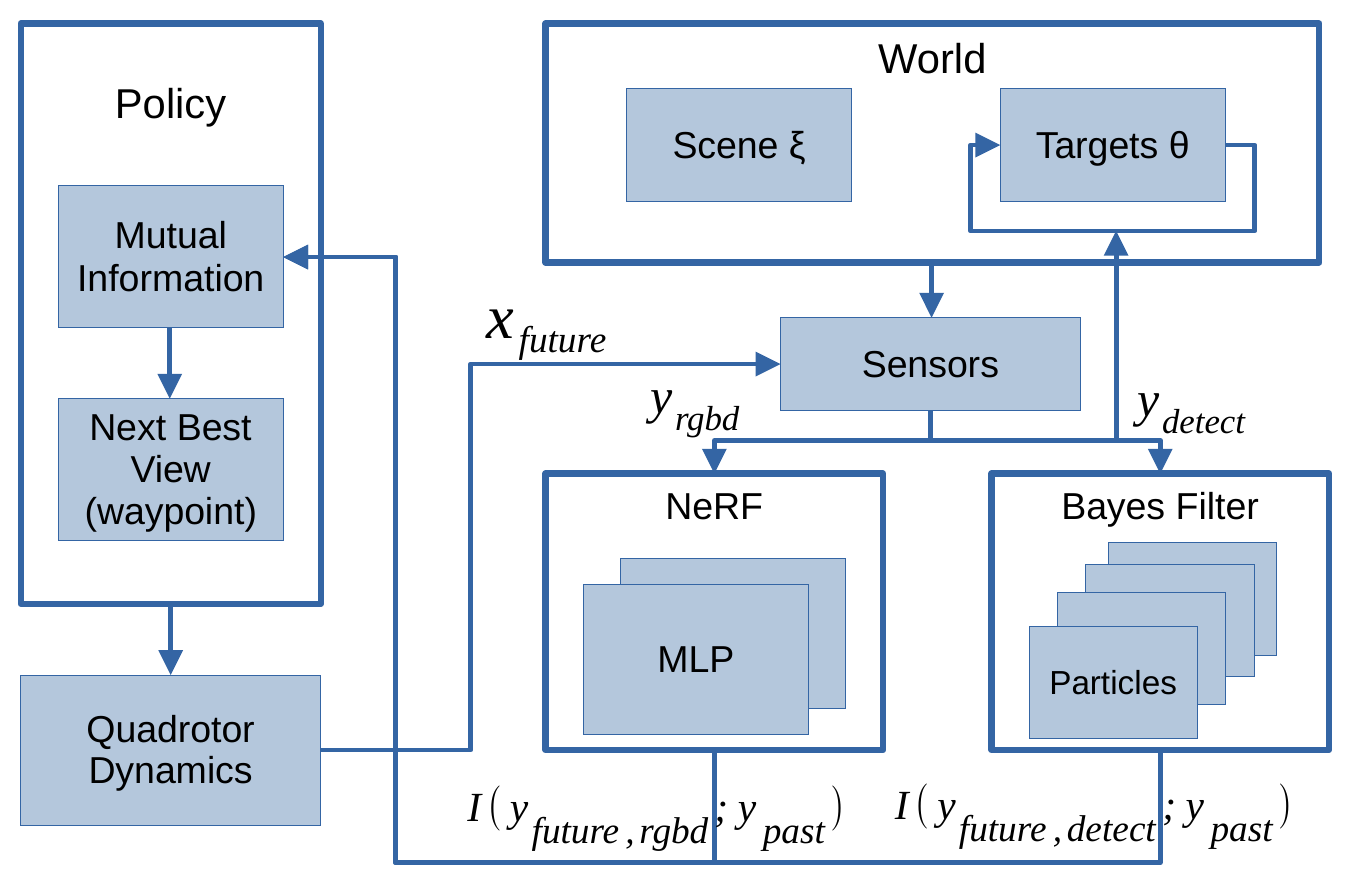}
    \caption{A schematic of the Active Scout system architecture.}
    \label{fig:schematic}
\end{figure}

We have three components to the observation $y$: images from the static scene $\xi$ and detections of the dynamic targets $\theta$. In this section, we will discuss first how to represent the scene when it is unknown to the scout via neural radiance fields (NeRFs). Next, we will describe how we use a Bayes filter to represent and maintain an estimate of target locations. Given these two representations, we will show how to calculate the most informative next location of the scout. Roughly speaking, the quadrotor first samples future observations $y_{\text{future}}$ from $p(y_{\text{future}} \mid x_{\text{future}}, \xi, \theta_t)$ and calculates the view $x_{\text{future}}$ that maximizes the information gain. At each step, it updates the Bayes filter and trains the NeRF using new observations $y_t \sim p(y_t \mid x_t, \xi, \theta_t)$. See the system architecture in the~\cref{fig:schematic} schematic. We will now focus on calculating
\[
    \I(y_{\text{future}};\ y_{\text{past}}).
\]
There are a few components to this: the mutual information corresponding to RGB and depth images and the mutual information corresponding to the target detections. These are discussed in the following sections.

\subsection{NeRF representation of the scene $\xi$}
If the map is unknown, the scout must first build a representation of the scene such that we can synthesize observations corresponding to future states $x_{\text{future}}$. We utilize the recent advances in NeRFs~\cite{li2023nerfacc, M_ller_2022} to build our representation. We will represent the scene $\xi: x \to (c,\sigma)$ as a neural radiance field which takes as input a pose $x \equiv (R,T) \in \text{SE(3)}$ and outputs color $c\in \reals^3$ and density $\sigma \in \reals_{+}$. Each location $x$ in the NeRF is parameterized using a positional encoding scheme called a multi-resolution hash map which feeds into a multi-layered perceptron (MLP) with 2 layers and 128 neurons per layer, see~\cite{M_ller_2022} for more details. We train the NeRF on the fly (using a few iterations of stochastic gradient descent after each time-step) using RGB images $y_{\text{rgb}}$ and depth images $y_{\text{depth}}$ with ground-truth pose from the quadrotor as it flies through the city.

The power of the NeRF representation is that it can be used to synthesize images from new viewpoints that the scout might not have seen before. The volume rendering equation lies at the heart of this capability; it is also useful to understand how the NeRF is trained. Assume a pinhole model for the camera where rays emanate from the focus at $T$. A point in the distance $d\in \reals$ from the focus along this ray has orientation $\delta R R$ which can be written as $x(d) = \delta R R (d,0,0)^\top + T$. The additional rotation $\delta R$ corresponds to all rays that lie in the field of view of the camera. The volume rendering equation samples points along this ray while querying the NeRF for color $c(x)$ and density $\s(x)$. The transmittance $p(d) = \exp \rbr{-\int_{d_0}^d \s(x(s)) \dd{s}}$ is the probability that a ray travels for an additional distance $d$ from the image image (which is at distance $d_0$) without encountering a solid object. Therefore, $p(d)\sigma(x(d))$ is the probability that the ray stops at $d$. Each rendered pixel has
\beq{\label{eq:rays}
    \aed{
        \text{color}: y_{\text{rgb}} &= \int_{d_0}^d \dd{s} p(s)\sigma(x(s))c(x(s)),\\
        \text{depth}: y_{\text{depth}} &= \int_{d_0}^d \dd{s} \sigma(x(s))s.
    }
}
These integrals are implemented using quadrature~\cite{mildenhall2020nerf} and techniques like~\cite{li2023nerfacc} can be used to speed up the rendering process by skipping known free space.

\begin{figure}[!htpb]
    \centering
    \includegraphics[width=0.32\linewidth]{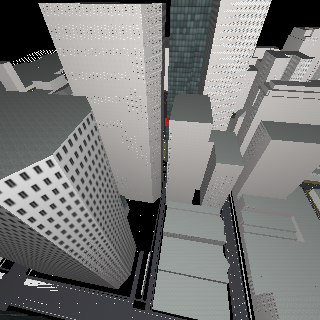}
    \includegraphics[width=0.32\linewidth]{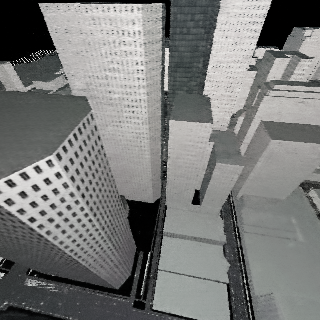}
    \includegraphics[width=0.32\linewidth]{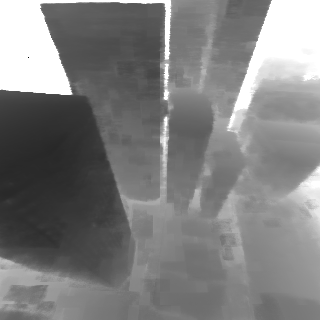}%
    \vspace*{0.25em}
    \includegraphics[width=0.32\linewidth]{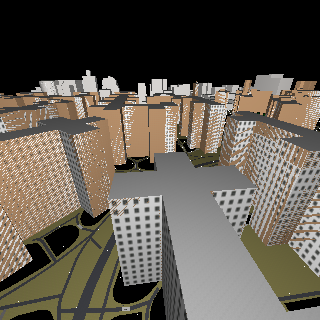}
    \includegraphics[width=0.32\linewidth]{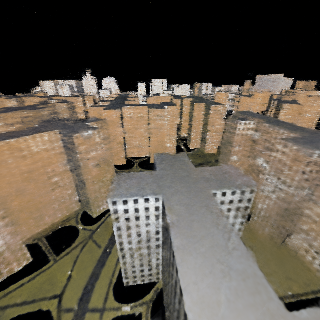}
    \includegraphics[width=0.32\linewidth]{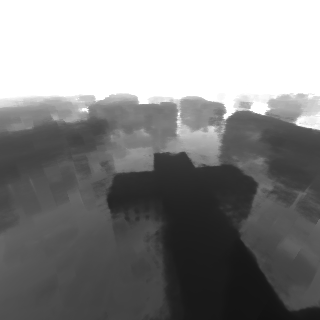}
    \caption{We show an example of two scenes: Philadelphia (top) and NYC StuyTown (bottom). For each scene we show the ground truth view RGB (left) compared against the NeRF rendering of RGB (middle) and depth (right). These RGB renderings have a PSNR $\approx 23$.
    }
    \label{fig:nerfreconstruct}
\end{figure}

We train the NeRF using images, depths, and their corresponding viewpoints collected online from a simulator that renders Open Street Maps data~\cite{OpenStreetMap}, see~\cref{fig:nerfreconstruct}. For each image and its viewpoint, we query the MLP for the color and density at different points along the ray. The rendered color and depth of each pixel are compared to their ground-truth values to calculate the loss $\ell = \lambda_1 \ell_{\text{rgb}} + \lambda_2 \ell_{\text{depth}}$; we use the $\ell_1$ loss for both terms. Stochastic gradient descent (SGD) is used to optimize the parameters of the MLP using this objective. We tuned the hyper-parameters $\lambda_1, \lambda_2$ such that the two terms have approximately equal magnitude during training. As the quadrotor flies through the scene, we expand the training dataset incrementally: adding new images and continuously performing SGD to update the NeRF. For each mini-batch, half the images are sampled from recent observations and the other half are sampled uniformly randomly from past observations~\cite{yu2023nerfbridge}.

We seek to calculate the mutual information $\I(y_{\text{future,rgb}}; y_{\text{past}})$ or $\I(y_{\text{future,depth}}; y_{\text{past}})$ for RGB and depth images respectively. The scene $\xi$ is a sufficient statistic of the past observations $y_\text{past}$. So we need to calculate $p(\xi \mid y_{\text{past}})$ the probability distribution over the unknown scene and $p(y_{\text{future,rgb}} \mid \xi)$ or $p(y_{\text{future,depth}} \mid \xi)$, the probability of scene given candidate future locations of the scout. In~\cite{he2023active}, we show that we can calculate a distribution over scenes using bootstrapped versions of the training dataset to build an ensemble of NeRFs that together represents $p(\xi \mid y_{\text{past}})$. We use two MLPs $\{\xi_k\}_{k=1}^2$ to set it to be $\delta_{\xi_1}(\xi)/2 + \delta_{\xi_2}(\xi)/2$, where $\delta$ denotes the Dirac delta distribution. NeRFs are not probabilistic models and therefore we cannot directly compute the likelihood of the scene. However, since the integrals of~\cref{eq:rays} are just an expectation, we can adjust them to calculate a variance for color: $\text{var}(y_{\text{rgb}}) = \int_{d_0}^d \dd{s} p(s) \sigma(x(s)) (c(x(s))-y_{\text{rgb}})^2$ and a similar expression for depth $\text{var}(y_\text{depth})$. If we assume that color and depth have a Gaussian distribution then we can calculate quantities like $p(y_{\text{future,rgb}} \mid \xi)$.

In unbounded scenes such as the city, some rays extend to infinity. This is problematic for our application because the scout may resort to exploring the sky overhead to maximize the information gain, i.e., the depth infinity, so there is always a mismatch between the volume density $\s$ predicted by the NeRF and the true volume density. We use a technique from~\cite{sünderhauf2022densityaware} to resolve this issue. The authors argue that most rays in the NeRF eventually hit some solid surface. One can therefore model the occupancy $y_{\text{occ}}$ along a ray as a Bernoulli random variable where the ray hits an obstacle with probability $1-p(d_{\max})$ and goes off to infinity with probability $p(d_{\max})$. For us, this is effectively an additional observation from the NeRF that depends on the volume density $\sigma$. We can calculate $p(y_{\text{occ}} \mid x_{t+\Delta t})$ which is probability over the Bernoulli distribution and also add an additional term to the mutual information objective. This term reduces the wasteful exploration that the scout performs to gain information about the open sky.


\subsection{Bayes Filter to estimate target locations}

We represent the probability distribution of the location of the $i^{\text{th}}$ ground target using $N$ particles
\[
    p(\theta^{(i)}_t \mid y_{\text{past}})  = \sum_{k=1}^N w_{t,k}^{(i)}\ \delta_{\tilde \theta^{(i)}_{t,k}}(\theta_t^{(i)})
\]
where each $\tilde \theta^{(i)}_{t,k} \in \reals^2$. In our problem, it is convenient to set up the particles densely on a fixed grid on the ground and set the weights of particles inside buildings (for a known map) to be zero. Updating the probability distribution after each time-step therefore corresponds to implementing a Bayes filter (rather than a particle filter). Note that the choice of the Bayes filter is merely for convenience; we could have also implemented a particle filter for this problem.

\begin{wrapfigure}{r}{0.4\linewidth}
    \centering
    \includegraphics[width=\linewidth]{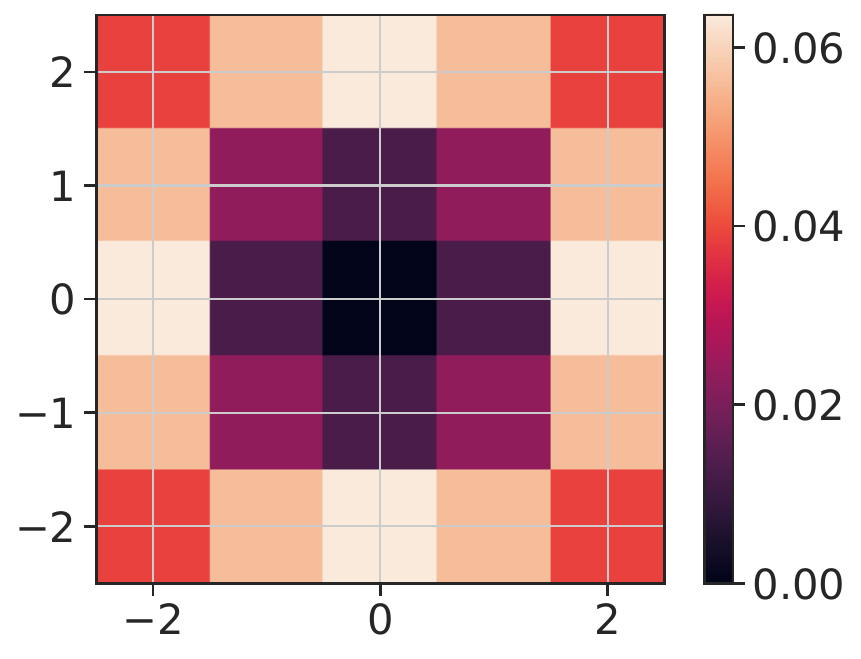}
    \caption{Target noise distribution model.}
    \label{fig:motion model}
\end{wrapfigure}
\paragraph{Motion Model} Depending upon the experimental setting, targets will either be stationary, or actively hide from the scout in the blind spots created by the buildings in the city. We compute the latter using a Dijkstra's algorithm for the target, this is described in the next section. In addition to this, we assume that the scout does not know the true motion model of the targets. If the target is located at the origin, the scout assumes that the probability of it moving to a nearby location is given by the probability distribution in the adjoining figure. It is important to choose this noise distribution carefully.\footnote{Consider the situation when the scout chooses a viewpoint that has line-of-sight of the particle with a high weight. If the noise distribution were Gaussian (symmetric around the origin), if the target moved away, and the scout did not obtain a detection, the posterior $p(\theta_t)$ would spread in all directions. In theory, this is not an issue, but in practice, such noise leads to a large variance in the posterior. This chosen noise distribution models a target that stays in the vicinity of the origin with a large probability but escapes in the direction of the four corners (as opposed to an arbitrary direction).}
The dynamics update for the Bayes filter corresponds to a convolution of the particle weights with the adjoining kernel, see~\cref{fig:motion model}.

\paragraph{Measurement Model}
Given a new observation $y_{\text{detect}}$ at time $t$ (target detections) we can update the Bayes filter as $p(\theta \mid y_{\text{past}}) \propto p(\theta \mid y_{0:t-1})\ p(y_t \mid \theta)$. We assume that when the target is in the field of view of the camera, the scout can detect it with a probability 0.95, i.e., with Bernoulli noise. When the target is hidden due to occlusions caused by the buildings, the scout cannot see the target. We assume that the scout can perfectly distinguish targets from each other and therefore the probability of incorrect data association is zero.

\paragraph{Calculating information gain for target tracking}
The posterior over the targets $p(\theta_t \mid y_{\text{past}})$ is a sufficient statistic of the past observations $y_\text{past}$ for detection. To calculate $\I(y_{\text{future,detect}};\ y_{\text{past}})$ we need to calculate the probability distribution of the target given some candidate future location of the scout. Given the ground-truth map, the statistic $p(\theta_t \mid y_{\text{past}})$, and under the assumption that the target is stationary, it is straightforward to calculate $p(y_{\text{future,detect}} \mid x_{\text{future}}, \theta_t)$; this is shown pictorially in~\cref{fig:1} (middle left). If we do not have the ground-truth map, we use the underlying voxel grid from the NeRF~\cite{li2023nerfacc,M_ller_2022} and the probability $p(y_{\text{occ}} \mid x_{\text{future}}, \xi)$ to trace a ray from the camera to each particle of the Bayes filter. If any voxel along this ray has a NeRF volume density above a threshold, the particle is unobservable from that pose. Whether we have the map or not, like we described above, the likelihood $p(y_{\text{future,detect}} \mid x_{\text{future}}, \theta_t)$ is a Bernoulli random variable with parameter 0.95. Therefore, we can calculate the information gain $\I(y_{\text{future,detect}};\ y_{\text{past}})$ for each target.\footnote{This calculation uses a simplistic dynamics model of the target (described above). This is a pragmatic choice. In principle, we could use a more complicated dynamics model, e.g., that targets hide in blind spots. But then calculating $p(y_{\text{future,detect}} \mid y_{\text{past}})$ is quite difficult; it would require us to run a different update using a particle filter for each putative scout location $x_{\text{future}}$.}

\subsection{Controlling the trajectories of the scout}
We use the standard differentially flat~\cite{Mellinger2011} dynamical model of a quadrotor where we are able to recover all other parts of the state and control inputs just from the four flat outputs: 3D Euclidean position and yaw. With a flat system, we can design trajectories that satisfy initial and final boundary conditions easily, e.g. any polynomial that fits these conditions, up to control constraints, is a dynamically feasible trajectory. Additionally, we include an independent fifth state, pitch, and altogether, waypoints for the quadrotor are in 5-dimensions.

At each time-step, we sample a set of putative future states $x_{\text{future}}$ in free space (straightforward with the ground-truth map, voxel grid underlying the NeRF is used otherwise). For each waypoint, we sample future observations $y_{\text{future}}$ to calculate mutual information. We make a key simplifying assumption:
\beq{
    \aed{
        \I(y_{\text{future}};\ y_{\text{past}})
        &= \I(y_{\text{future,rgb}};\ y_{\text{past}}) + \I(y_{\text{future,depth}};\ y_{\text{past}})\\
        &+ \l \sum_{i=1}^m \I(y_{\text{future,detect}}^{(i)};\ y_{\text{past}}).\\
    }
    \label{eq:nerf_decomposition}
}
Here, the first term corresponds to the information gain for the scene (as if there were no targets, split between independent terms for RGB, depth, and occupancy) and the second term corresponds to information gain for the targets (as if the scene were known). Note that the probability density of the target locations in our Bayes filter is certainly a function of the scene (e.g., observations respect occlusions, targets cannot enter buildings, etc.). This decomposition allows us to calculate mutual information without worrying about calculating the joint distribution of the scene and targets. The hyper-parameter $\l=10$ enables the scout to trade-off between target tracking and learning the scene (which helps target tracking in the long-term even if it forgoes near-term tracking performance).

We calculate 10 different scout distributions $\varphi \equiv p(x_{\text{future}} \mid y_{\text{past}})$ for~\cref{eq:MIobj}; each of these distributions is represented by 10 particles centered around some waypoint in 3D space. Calculating the mutual information objective is an expensive calculation because it involves many different queries of the NeRF; depending upon the application one could use fewer waypoints. Instead of an $\argmax$ in~\cref{eq:MIobj} we experimented with a more stochastic policy where waypoints are chosen using multinomial sampling; those with larger $\I$ are more likely to be chosen. This scheme breaks the greedy formulation that can cause the scout to be stuck in local minima. Scout trajectories between successive waypoints $x_t \to x_{t+\Delta t}$ are calculated using Dijkstra's algorithm combined with rotorpy~\cite{folk2023rotorpy} to solve an quadratic optimization problem that parameterizes the flat outputs using a $7^{th}$ order polynomial to minimize the integral of the squared snap. Since the camera pitch is independent of the quadrotor dynamics, we linearly interpolate the pitch along this trajectory. We found it helpful to perform an additional $2\pi$ yaw rotation with some modulation of the pitch at the end of the trajectory; this gives the scout extra information to train the NeRF as well as a larger potential for spotting targets.

\section{Simulation Experiments}
\label{s:experiments}

\begin{wrapfigure}{r}{0.5\linewidth}
    \centering
    \includegraphics[width=0.99\linewidth]{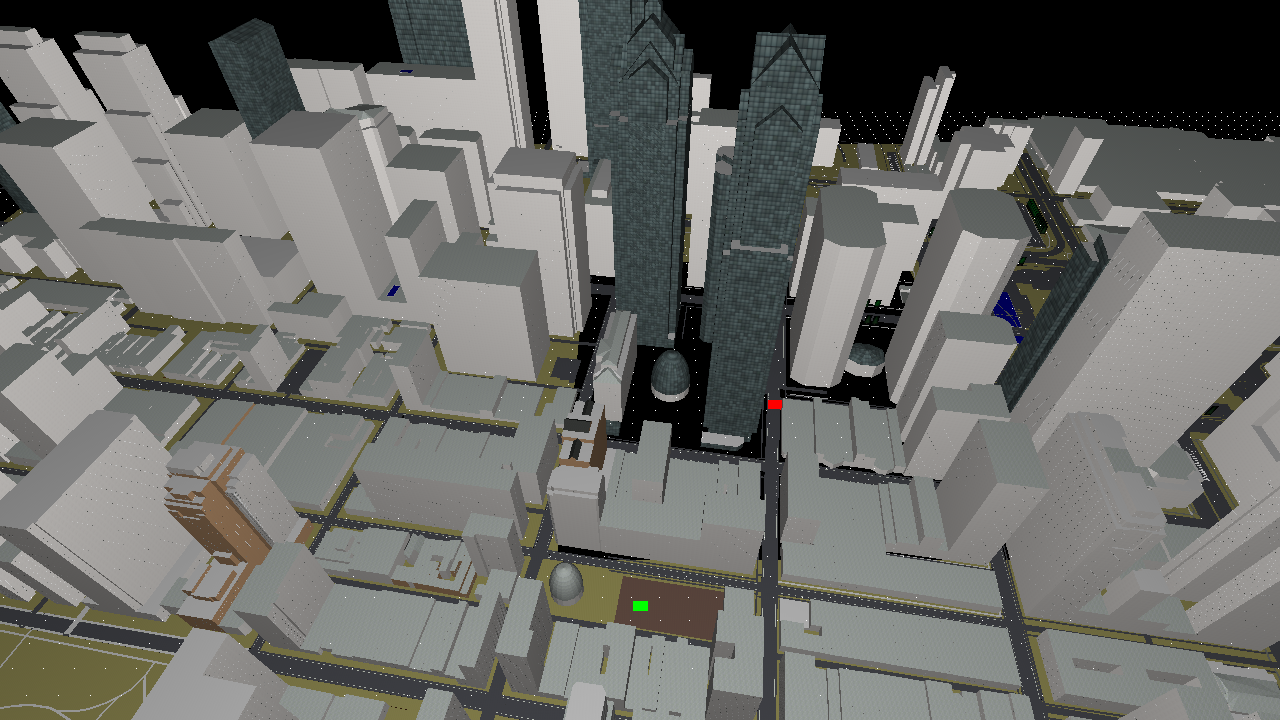}%
    \vspace*{0.25em}
    \includegraphics[width=0.99\linewidth]{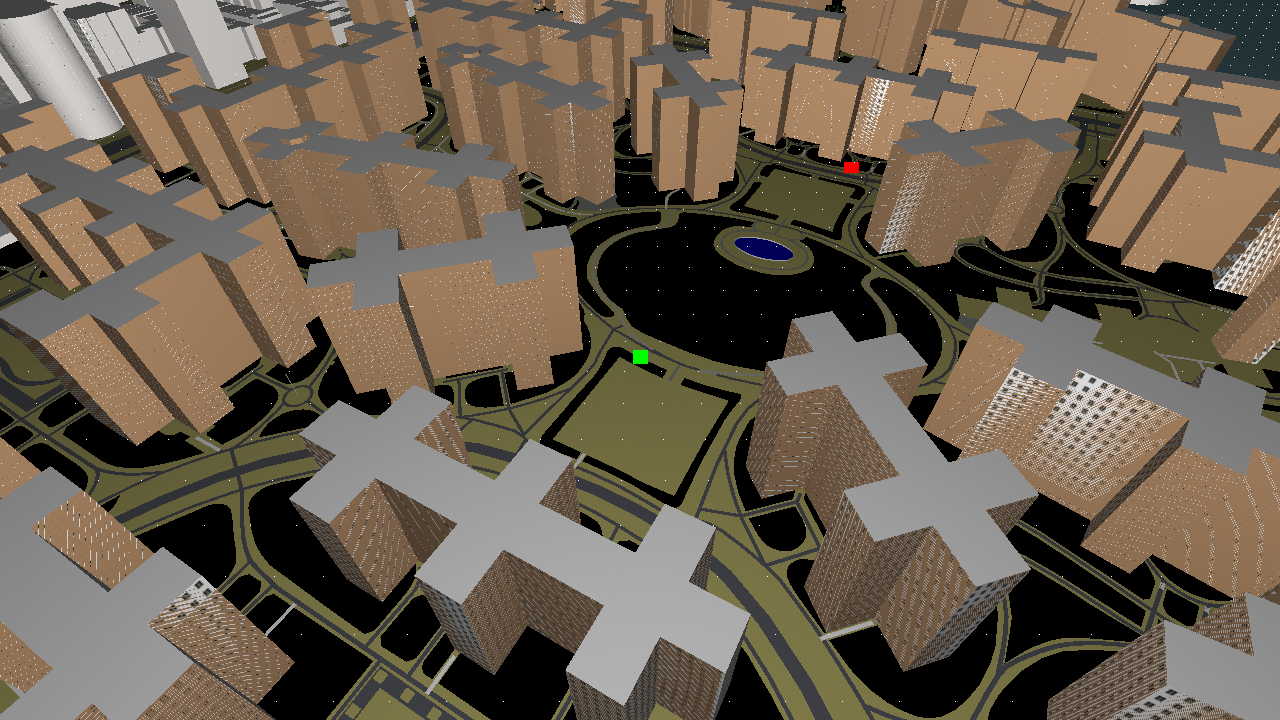}
    \caption{A preview of the Philadelphia (top) and NYC StuyTown (bottom) scene in third person view with a green and a red target in view.}
    \label{fig:cityview}
\end{wrapfigure}

We built a simulator using Open Street Maps~\cite{OpenStreetMap} that can render the scene (buildings, locations of the targets etc.) using OpenGL. We focus on two specific maps for our simulation experiments. The first is of Center City Philadelphia, see~\cref{fig:1} and~\cref{fig:cityview} (top), where we set the coordinate origin to be at latitude: 39.9517 and longitude: -75.1671. The map is normalized such that each unit in each direction is roughly 1m. Similarly, our second map is of the apartment complex StuyTown in NYC, see~\cref{fig:nerfreconstruct} (bottom) and~\cref{fig:cityview} (bottom), with the coordinate origin set to be at latitude: 40.7327 and longitude: -73.9771. The Center City Philadelphia map has taller buildings with irregular heights creating more intricate occlusions than the NYC map that has shorter buildings and more space in between. We bound the altitude the scout can travel such that in the Philadelphia scene it can only fly up to an altitude of 150m whereas in the NYC scene it can fly up to 100m. We do this because if the scout is able to fly to an unbounded height, it is easy to observe the entire map from a single vantage point. From this simulator, given a pose $\in \text{SE(3)}$ we are given RGB and depth images.

The scout has a camera with a field of view of 90 degrees and receives RGB and depth images of size 320$\times$320. Each experiment begins with the scout at the origin and collecting observations by first increasing the altitude to some designated maximum; it then performs a $2\pi$ yaw rotation with some random perturbations to the pitch to fit an initial model of the scene with 30 images, 1 image per step. We train the NeRF on these initial images for 4,000 training iterations before the experiment begins, and train for 4,000 more iterations after each waypoint is reached. Targets are randomly initialized on the ground plane in free space.

Our figure of merit is the mean squared error (RMSE) between the scout's estimate of the target (mean of the posterior in Bayes filter) and the target's true position. To highlight targets that are currently being observed, we plot the minimum RMSE as a high opacity plot with the color indicating which target is contributing to this value. In comparison, we plot the maximum RMSE as a low opacity plot and the corresponding target color to represent the neglected target. We also summarize these quantities in~\cref{tab:quant}.

\subsection{Scout Policies}

We evaluate variations of scout policies over 2 target policies (stationary and active) in 2 scenes. Each experiment will begin with the scout executing an initialization phase for which they will fly to their maximum height and do a $2\pi$ yaw rotation scan of the scene. After that, the scout will perform 40 planning steps with each step having 30 control steps. The 4 methods below describe the scout's policies and information strategy. We will have:
\begin{itemize}
    \item \textbf{GTmap+MAP:} ground truth map + greedy follower (MAP: Maximum A Posteriori), 
    \item \textbf{GTmap+MI:} ground truth map + mutual information,
    \item and \textbf{NeRF+MI:} NeRF + mutual information.
\end{itemize}

The following experiments will evaluate whether neural radiance fields (NeRF) are a valid replacement for a ground truth map (GTmap) given some control policy utilizing the map and target representations. In the NeRF+MI experiments, we will train a NeRF from data collected on the fly and execute mutual information (MI) based control from that representation,~\cref{eq:nerf_decomposition}, and MI over the Bayes filter. In comparison, GTmap+MI will use the ground truth map and MI calculated over the Bayes filter, i.e. $I = \sum_{i=1}^m \I(y_{\text{future,detect}}^{(i)};\ y_{\text{past}})$.

As our control baselines, we will use the ground truth map plus a greedy control policy that takes control actions that maximize a posteriori (MAP) over the Bayes filter. The MAP policy can be thought as choosing the pose that gives the maximum expected value over the detected target locations given the future poses such that
\[
 u_{\text{future}} \in \argmax_{\varphi} \mathbb{E}_{\theta} [y_{\text{future,detect}}^{(i)}; y_{\text{past}}|x_{\text{future}}],
\]
where $\varphi \equiv p(x_{\text{future}} \mid y_{\text{past}})$.

\subsection{Target Policies}
For each of the scout policies as described above, we evaluate them on 2 different types of control polices: stationary targets and active targets. Stationary targets will be our exploration baseline for these methods. On the other hand, active targets are dynamic and are deliberate in the locations they select to go to as they are aware of the scout's actions and views. 

\subsubsection{Stationary Targets (exploration task)}

\begin{figure}[!htpb]
    \centering
    \includegraphics[width=0.32\linewidth]{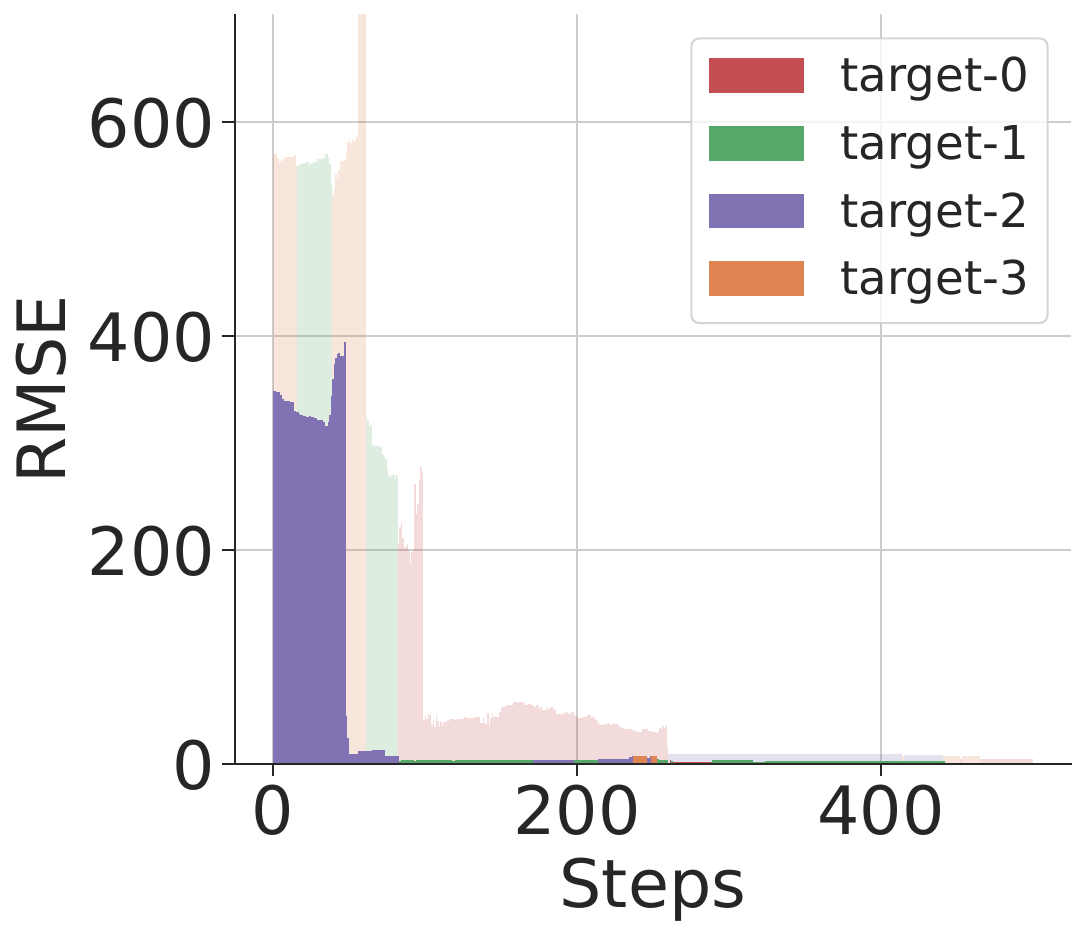}
    \includegraphics[width=0.32\linewidth]{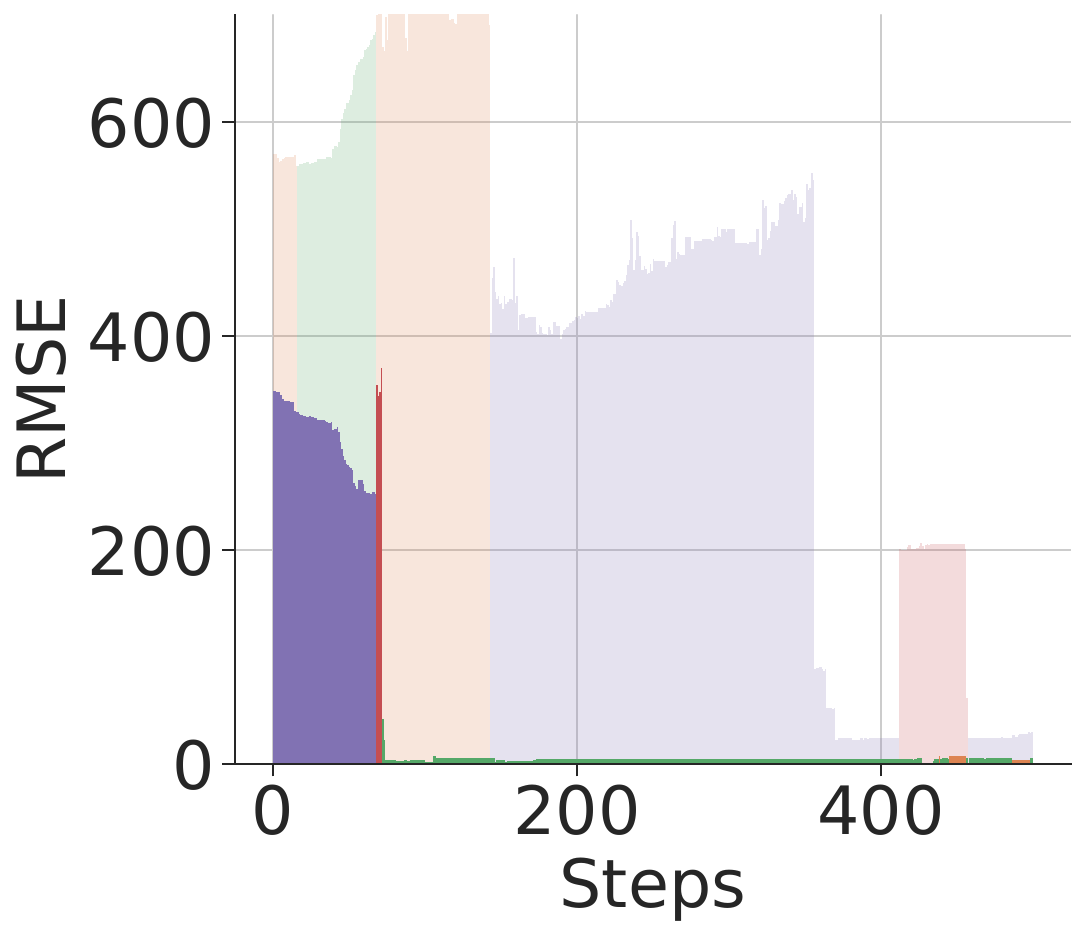}
    \includegraphics[width=0.32\linewidth]{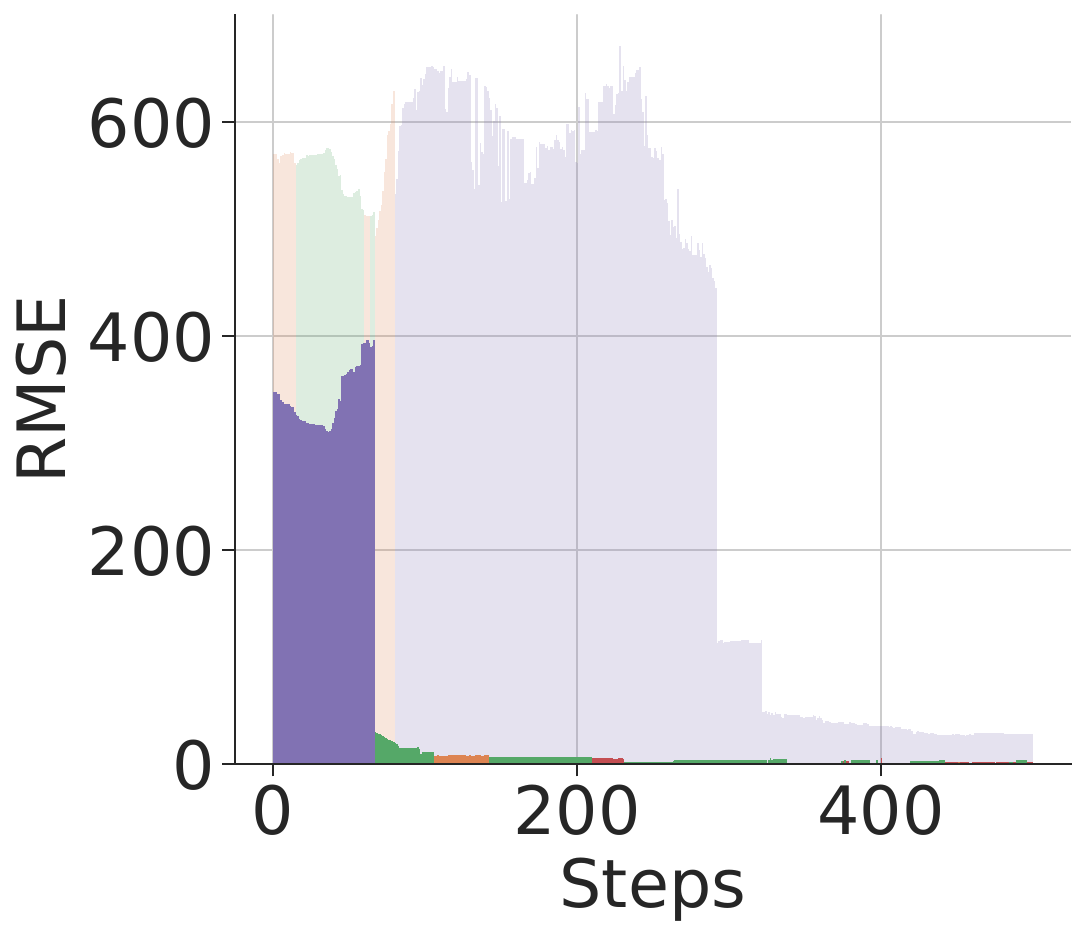}
    \includegraphics[width=0.45\linewidth]{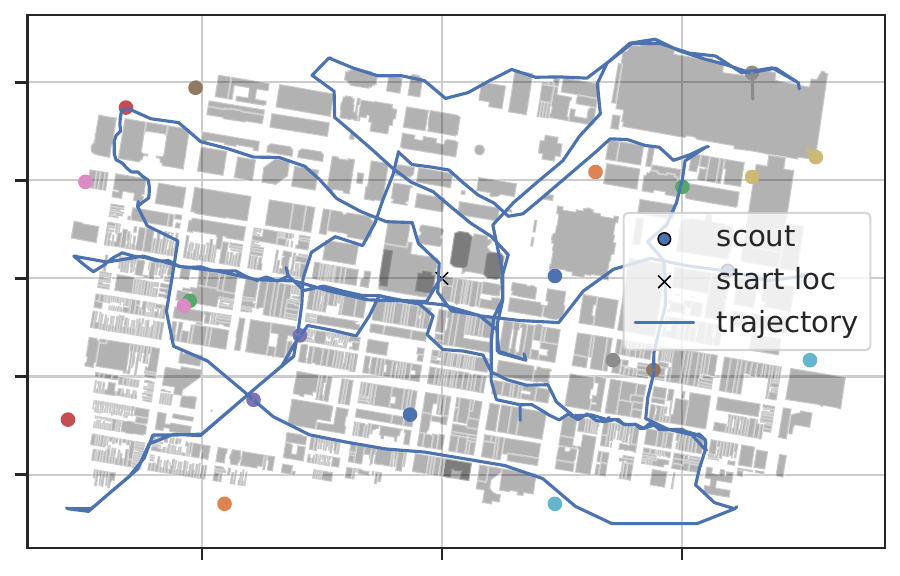}
    \includegraphics[width=0.45\linewidth]{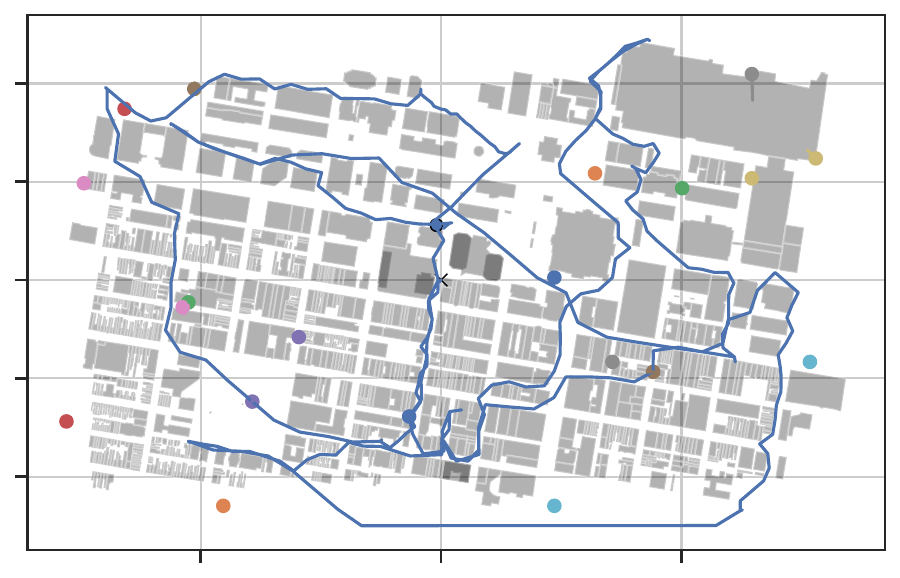}
    \caption{\textbf{Top:} In the Philadelphia scene, we test the three methods GTmap+MAP (left), GTmap+MI (middle), and NeRF:4k+MI (right) on \textbf{20 stationary} targets randomly initialized within the scene (the legend shows an example of the first 4 targets colors) and show the RMSE plot over control steps. \textbf{Bottom:} We show the scout's trajectory (blue line) using the GTmap+MAP (left) and NeRF:4k+MI (right) and the locations of the 20 targets in the 2D plot of Philadelphia.}
    \label{fig:stationary}
\end{figure}

In the first experiment, see~\cref{fig:stationary}, we test the three methods against 20 stationary targets in the Philadelphia scene. We observe that all methods are able to localize all the targets within some time. Given a uniform prior over target locations, it can be seen that GTmap+MAP is the quickest at finding all the targets. MI based policies take longer to find all of the targets but they seem to be more thorough in exploring the map. For the observant reader, they will notice that the RMSE increases for some targets for some time. This happens because we still apply the motion model to bring about some uncertainty in the filter. We will see later that when the targets are dynamic, the greedy policy is suboptimal. 

This experiment also shows that our method does not need to know how many targets are in the scene a priori since we have a camera sensor that can provide good target identification. In all our experiments, we add a new filter for each new target $i$ that is found such that there is always 1 extra filter that has not been assigned a target. By always maintaining one extra filter, the scout always has a small opportunity to select a view that promotes exploration. When a new target is found the extra filter is assigned and a new one with a uniform prior is created. We see that in~\cref{fig:stationary} for each method, all targets have been identified and the filter for each target has converged to the true locations.

\subsubsection{Actively hiding targets}
\label{ss:activetargets}
Targets with the active policy have knowledge of the scout and maintain a history of graph nodes that the scout has seen during online execution, i.e they have access to $y_{\text{detect}}$. During each iteration, an active target will randomly select a graph node from the list of particles the scout has not seen and move there via a path found by Dijkstra's. We observe that targets that follow this policy will actively hide behind buildings, i.e. more occluded locations, locations that are harder for the scout to see. This is because over time as the scout sees more parts of the map, the parts that have not been seen must be places of the map that the scout must be more deliberate to be able to view. We reset this observed particles buffer at regular intervals (every 10 planning steps) so that targets can return to old graph nodes. This type of reset forces the scout to return to previously seen locations for the sole reason of observing the targets.

In~\cref{tab:quant}, we summarize and provide a teaser of the experiments that will be discussed in this section. We measure the root mean squared error (RMSE) in meters (m) of the targets' true positions compared to the estimated positions. In the table we report the tracking error mean, minimum, maximum, and corresponding standard deviation which is computed by averaging the RMSE of each target over all targets, over the entire length of the experiment, and over 3 arbitrary seeds (72, 80, 88). 

\begin{table}[!htpb]
    \centering
        \resizebox{\linewidth}{!}{
        \renewcommand{\arraystretch}{1.25}
        \begin{tabular}{l c r r r}
        \toprule
        City & Method & Tracking Error (TE) Mean (m) & TE Min (m) & TE Max (m) \\
        \midrule
        Philly & GTmap+MAP & 190.668 $\pm$ 46.439 & 1.326 $\pm$ 0.802 & 603.155 $\pm$ 33.963\\
        Philly & GTmap+MI & 124.671 $\pm$ 20.088 & 0.583 $\pm$ 0.283 & 525.107 $\pm$ 92.670\\
        Philly & NeRF:4k+MI & 133.818 $\pm$ 24.080 & 0.893 $\pm$ 0.293 & 563.084 $\pm$ 61.766\\
        Philly & NeRF:2k+MI & 153.711 $\pm$ 29.144 & 1.125 $\pm$ 0.423 & 547.610 $\pm$ 90.076\\
        Philly & offlineNeRF+MI & 117.800 $\pm$ 21.931 & 1.013 $\pm$ 0.569 & 550.639 $\pm$ 16.408\\
        \midrule
        NYC & GTmap+MAP & 163.216 $\pm$ 18.319 & 0.677 $\pm$ 0.204 & 572.044 $\pm$ 38.461\\
        NYC & GTmap+MI & 137.121 $\pm$ 12.754 & 0.976 $\pm$ 0.263 & 550.491 $\pm$ 50.295\\
        NYC & NeRF:4k+MI & 145.057 $\pm$ 34.356 & 1.492 $\pm$ 1.252 & 586.810 $\pm$ 39.409\\
        NYC & NeRF:2k+MI & 144.363 $\pm$ 29.417 & 1.573 $\pm$ 0.934 & 594.730 $\pm$ 49.646\\
        NYC & offlineNeRF+MI & 167.537 $\pm$ 24.989 & 0.781 $\pm$ 0.330 & 570.791 $\pm$ 65.133\\
        \bottomrule
        \end{tabular}
        }
\caption{We provide a table of experiments that will be discussed in~\cref{ss:activetargets} where the root mean squared error (RMSE) tracking error is averaged over the 4 targets and 3 seeds across the trajectory. We observe that NeRF based policies are similarly performant to ground truth but GTmap+MI is the most performant.}
    \label{tab:quant}
\end{table}

We find that GTmap+MI is the most performant and outperforms GTmap+MAP, a very greedy policy that does not do a good job switching between targets to minimize the overall tracking error. We also observe that NeRF-based policies and their variants with mutual information perform admirably against their ground truth counterparts. In this paper we provide 3 variants to the NeRF+MI policy including NeRF:4k, NeRF:2k, and an offline NeRF. NeRF:4k+MI is the standard setup for which we train the NeRF for 4,000 training steps in between each planning step. NeRF:2k+MI is similar however it is only trained for 2,000 steps for every planning step. Finally, offlineNeRF+MI is an experiment in which the NeRF is \textit{pretrained} on the scene from images collected with the scout following mutual information for 40 planning steps each with 4,000 training steps. When that phase in complete, the experiment begins and follows the procedure of the two previous NeRF+MI policies where it continues to train the NeRF online but for only 2,000 training steps per planning step. From the metrics in~\cref{tab:quant}, we observe that more training steps generally leads to better tracking performance. It is important to note that although the experiments are quite stochastic we still see these intuitive trends emerge. We surmise that with more runs, GTmap+MI will overall continue to be the most performant and offlineNeRF+MI will be close behind. These observations leads to our conclusion that NeRFs are a valid replacement to ground truth maps.

\begin{figure}[!htpb]
    \centering
    \includegraphics[width=0.32\linewidth]{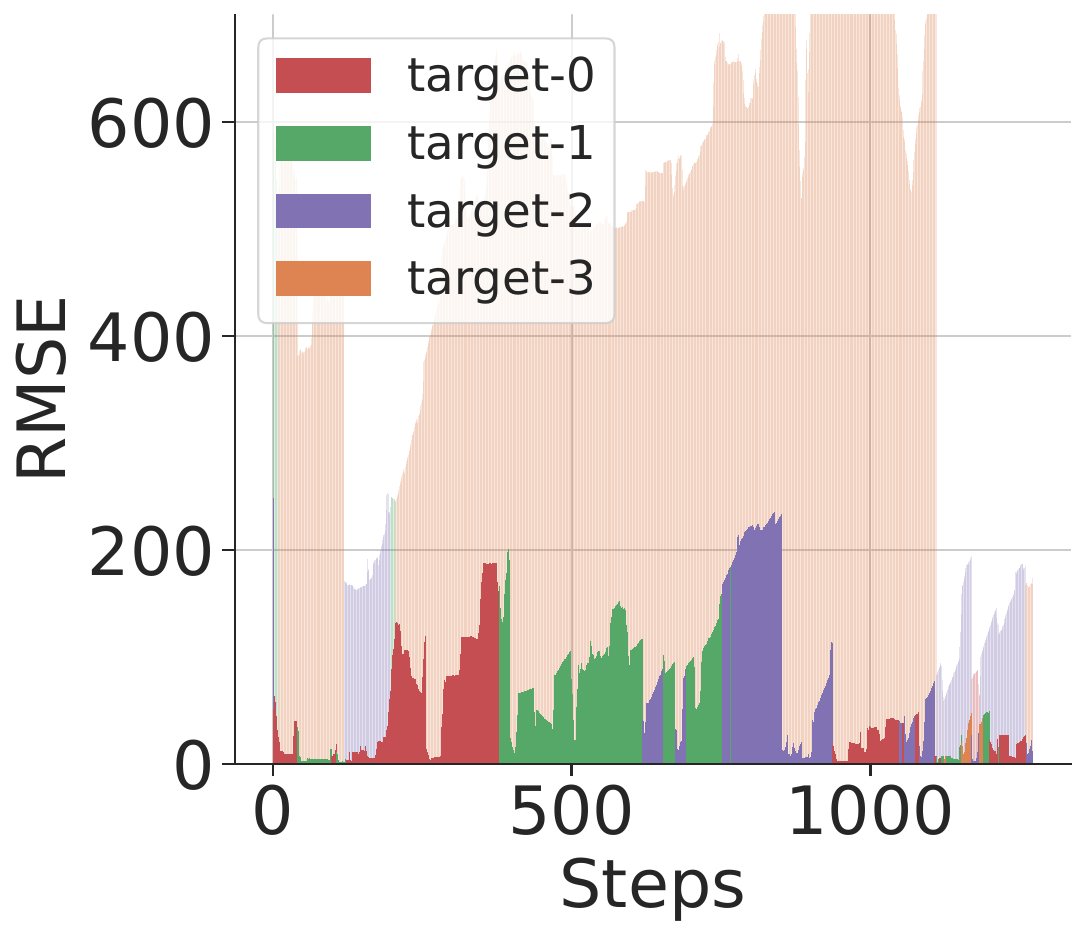}
    \includegraphics[width=0.32\linewidth]{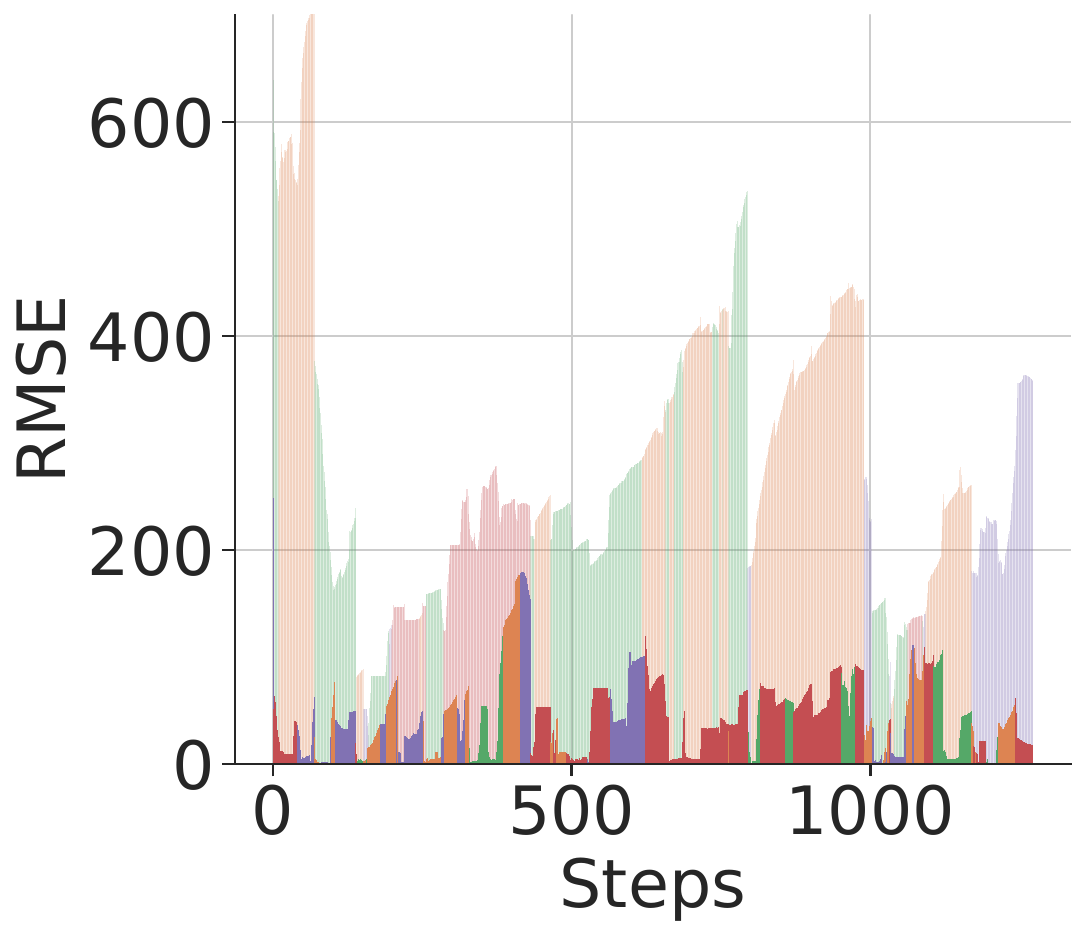}
    \includegraphics[width=0.32\linewidth]{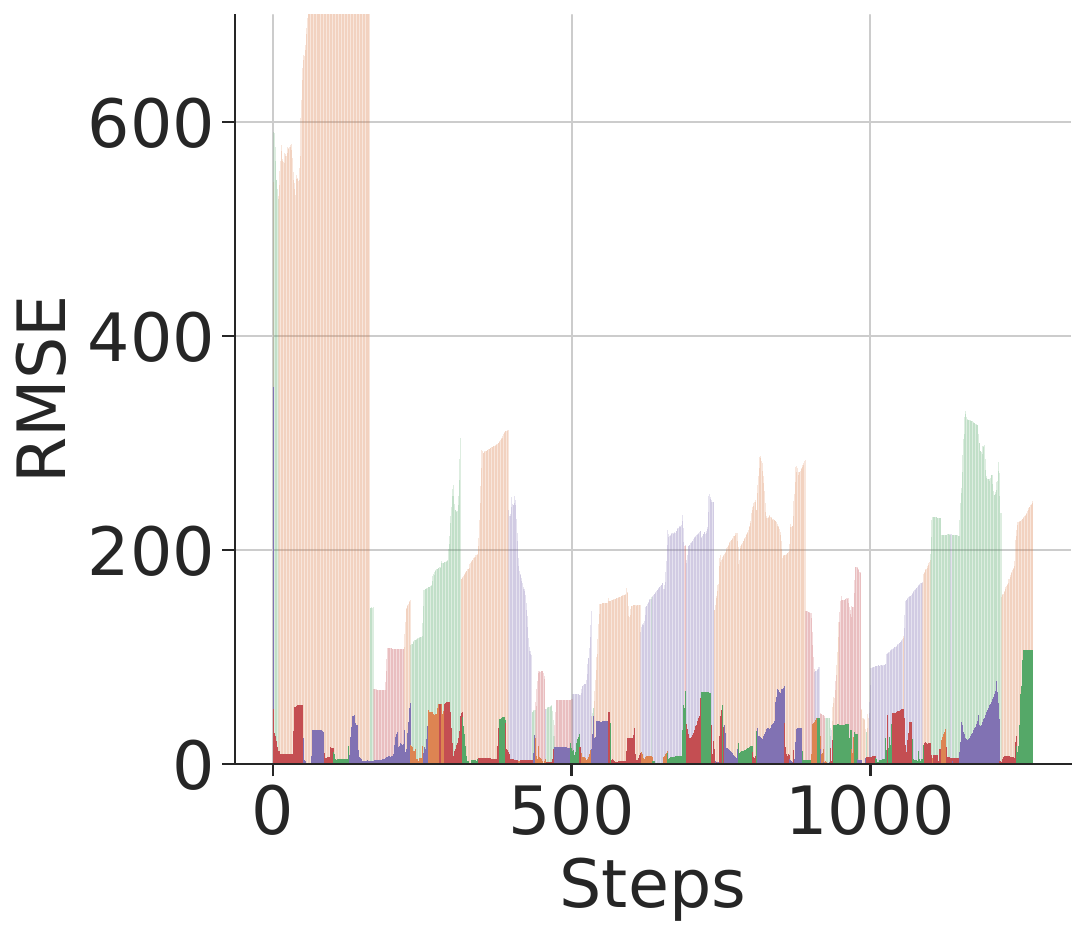}\\
    \includegraphics[width=0.32\linewidth]{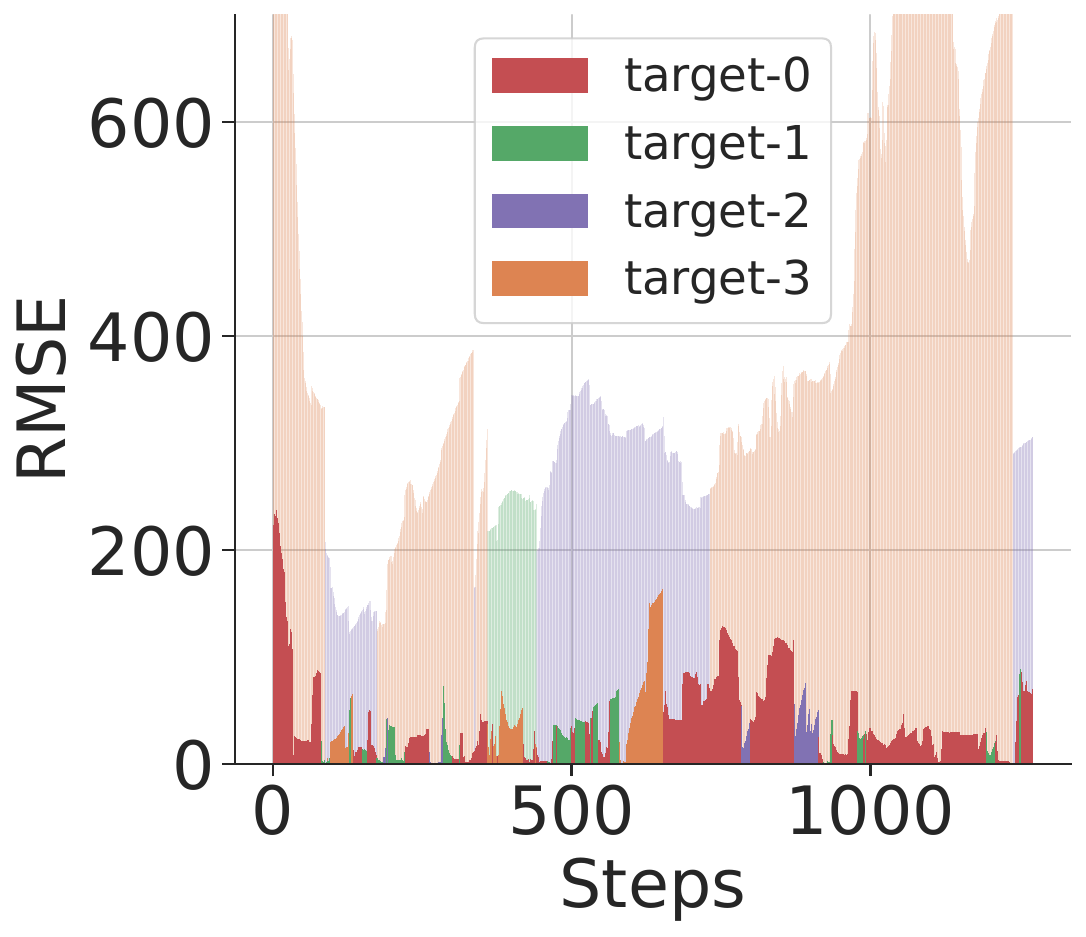}
    \includegraphics[width=0.32\linewidth]{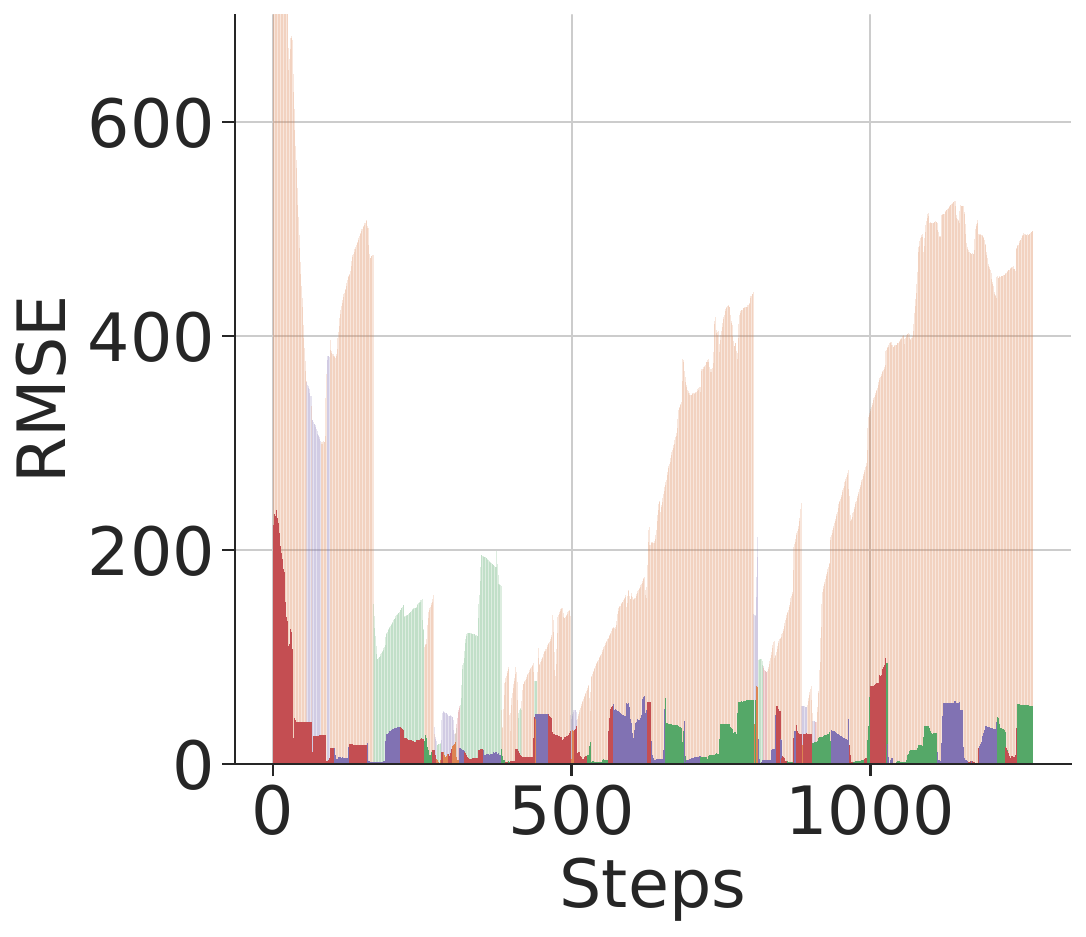}
    \includegraphics[width=0.32\linewidth]{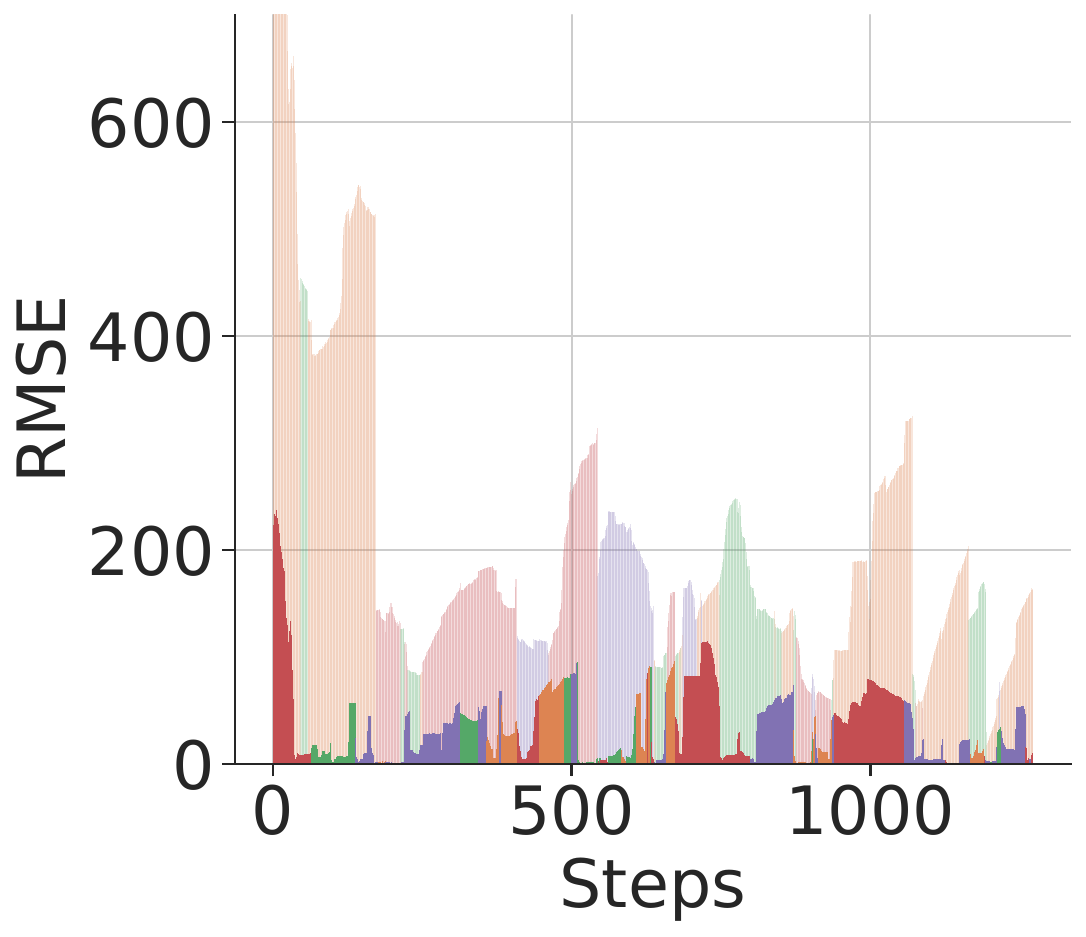}
    \caption{In these plots the 4 targets follow the \textbf{active} policy:  targets move to locations on the map that the scout has not seen. \textbf{Top:} In the \textbf{Philadelphia} scene, we compare 3 types of scout policies (left: GTmap+MAP, middle: GTmap+MI, right: NeRF:4k+MI). The plots display maximum (light opacity) and minimum (full opacity) mean squared error (RMSE) for the estimated target location against the true target position (seed 88). The color plotted shows which target is contributing to the maximum or minimum error. 
    \textbf{Bottom:} Similar to the plots in the top row, the bottom row of plots are the 3 scout policies in the \textbf{NYC StuyTown} scene tracking active targets.
    }
    \label{fig:RMSE_active}
\end{figure}

In~\cref{fig:RMSE_active}, we plot the RMSE of the scout's estimate of the target over control steps in the Philadelphia scene (top row) and the NYC StuyTown scene (bottom row). After the initialization phase (first 30 control steps) in both scenes, there is a low minimum RMSE as at least 1 target has been seen. This is indicated by the high opacity plot and the color denotes which target contributes to that value, i.e. the red target has the smallest RMSE of the 4 and is plotted. On the other hand, the orange target has the largest RMSE and is plotted in orange with light opacity. In both scenes and all methods, by approximately step 200, all targets have been spotted at least once. From there on out the targets are more actively hiding and is the cause of error for the scout for the rest of the episode. We observe that the NeRF:4k+MI (right) experiments have similar trends to that of GTmap+MI (middle) where different targets are observed over time (varying colors of the full opacity plots) while allowing some targets to remain undetected for some time (light opacity plots). We can see with the variation in colors that over time as targets have not been observed for some time they will gain in RMSE, but upon observation that RMSE will become small and a different target will contribute to the large RMSE.

Next, compare the MI based policies to the MAP based policies in~\cref{fig:RMSE_active}. In the Philadelphia scene (top), the greedy MAP agent with the ground truth map (left), does a poor job of observing the orange target, allowing its RMSE to explode. After some time it does find the target and ends the episode with smaller RMSE. The MAP greedy scout in both scenes can be seen to miss a target for quite a while. We attribute this to the scout greedily checking locations with a large posterior, i.e. places the filter indicates the target is likely to be. Since the targets move to places that are hard to find for the scout, the greedy scout does a bad job of checking hard to view areas, therefore missing hard to find targets.

\begin{figure}
    \centering
    \includegraphics[width=0.32\linewidth]{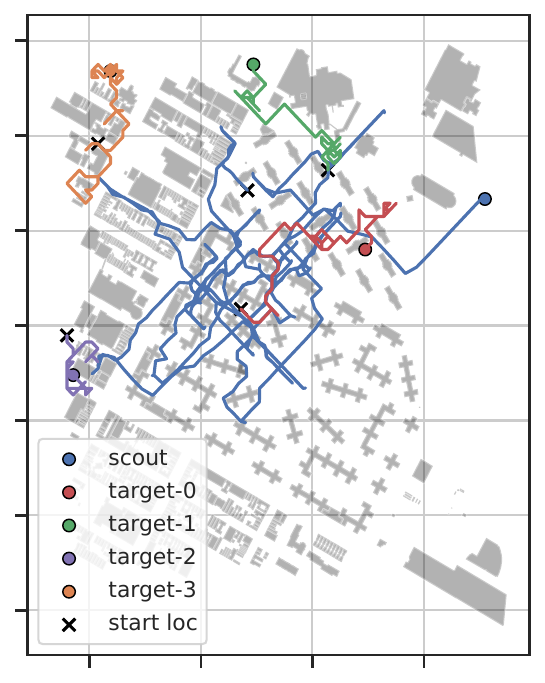}
    \includegraphics[width=0.32\linewidth]{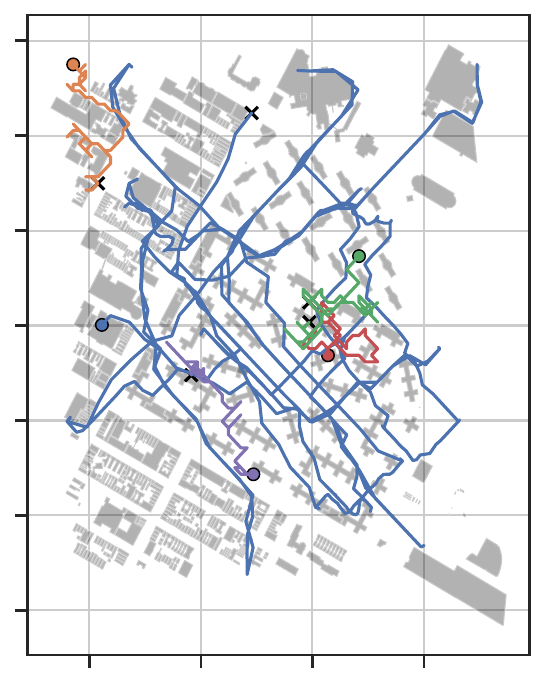}
    \includegraphics[width=0.32\linewidth]{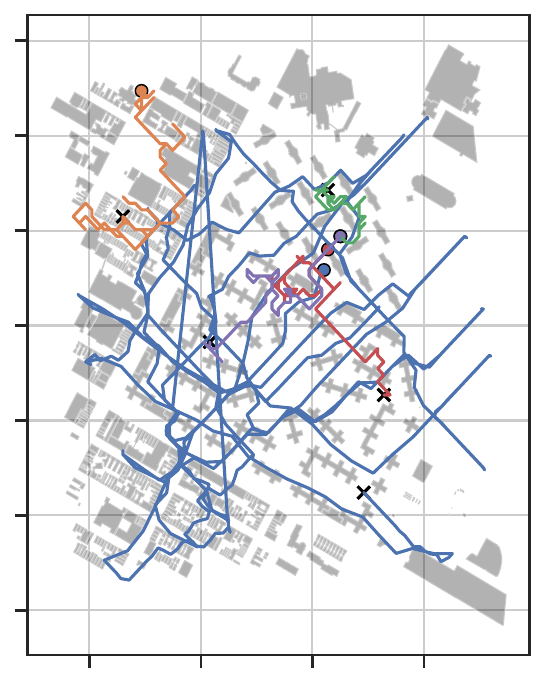}
    \caption{We show the active targets' (colored lines) trajectories in the NYC StuyTown map so we can observe the scout's trajectory (blue line) executing the GTmap+MAP (left), GTmap+MI (middle), and NeRF:4k+MI (right) policies. The trajectories plotted start at step 500 (black `x') until the end of the episode (circle).}
    \label{fig:traj-nyc}
\end{figure}

We want to note an interesting observation in the NYC StuyTown scene with respect to the \textbf{orange} target as it produces large RMSE for the GTmap+MAP and the GTmap+MI scout, see~\cref{fig:RMSE_active} bottom row. Looking at~\cref{fig:traj-nyc}, we surmise that adding in the photometric and geometry information, i.e. $y_{\text{future,rgb}}$ and $y_{\text{future,depth}}$, encourages the scout to do a more thorough exploration of the map that leads to finding the orange target in comparison to just the MI or MAP from the filter. Compare the blue trajectories of the scout. On the left is the NeRF+MAP scout that greedily rotates between the estimated target locations. It seems to miss the orange target as it moves to the top left most part of the map. In comparison, MI based policies (middle and right) do a better job in exploring the map even when targets are in view and in the vicinity. Even so, GTmap+MI only does a single pass to the top left of the map resulting in a large RMSE in~\cref{fig:RMSE_active} (bottom middle). NeRF:4k+MI (right) does the best at exploring with multiple passes into the top left portion of the map and this results in the scout maintaining a low RMSE for the orange agent as seen in~\cref{fig:RMSE_active} (bottom right). 

\begin{figure}[!htpb]
    \centering
    \includegraphics[width=0.45\linewidth]{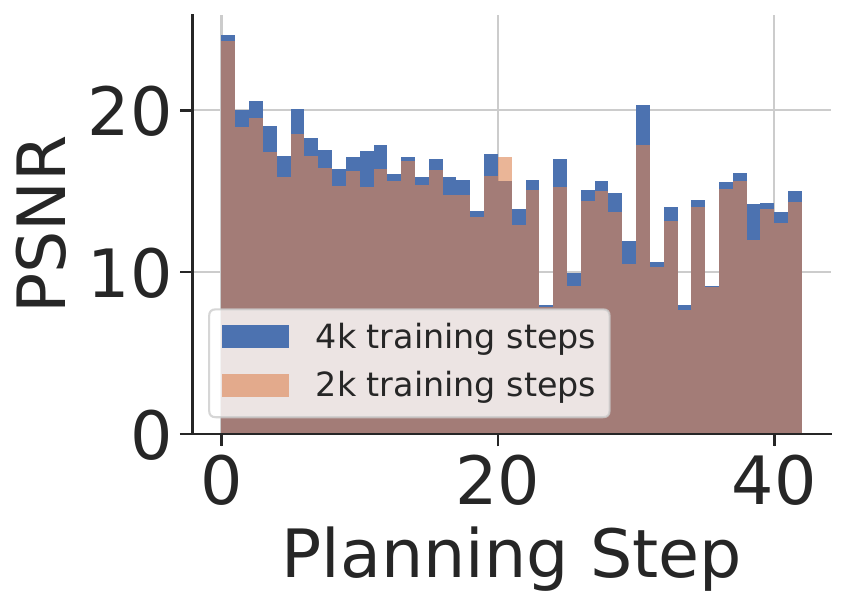}
    \includegraphics[width=0.45\linewidth]{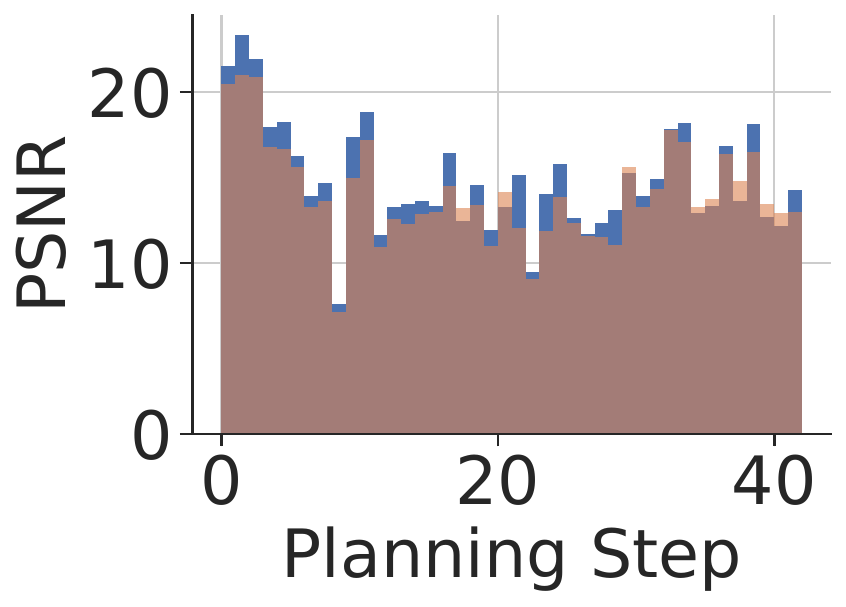}
    \caption{During each planning step after a set of control steps, the scout takes some time train the NeRF. We plot the average peak signal to noise ratio (PSNR) between the test set (20 most recently collected images) and the ground truth after 2,000 training steps and 4,000 training steps for the Philadelphia (left) and NYC (right) scene. A larger PNSR is better.}
    \label{fig:psnr}
\end{figure}

Finally, we take a look at the reconstruction quality of the NeRF trained on the fly in~\cref{fig:psnr}.  We plot the peak signal to noise ratio (PSNR) which is a metric to describe the reconstruction quality of the NeRF image against the ground truth image. In~\cref{fig:psnr}, after each set of control steps where the scout is collecting images, it stops to train the NeRF for a set number of training steps. We report the average PSNR of the most recent 20 images collected prior. We observe that as the scout selects new locations to travel to, it expands the scene for which the NeRF must learn to reconstruct. As the dataset grows and the scene grows, it becomes more difficult for the NeRF to render high quality images: due to model capacity or data insufficiency. Although this is the case, we have seen previously that the scout is still able to do a good job in tracking the targets. 
We plot the PSNR evaluation at step 2,000 and at step 4,000. In past experiments, e.g.~\cref{fig:RMSE_active} (right), the 4,000 step result is what the scout utilized during the episode.

\begin{figure}[!htpb]
    \centering 
    \includegraphics[width=0.35\linewidth]{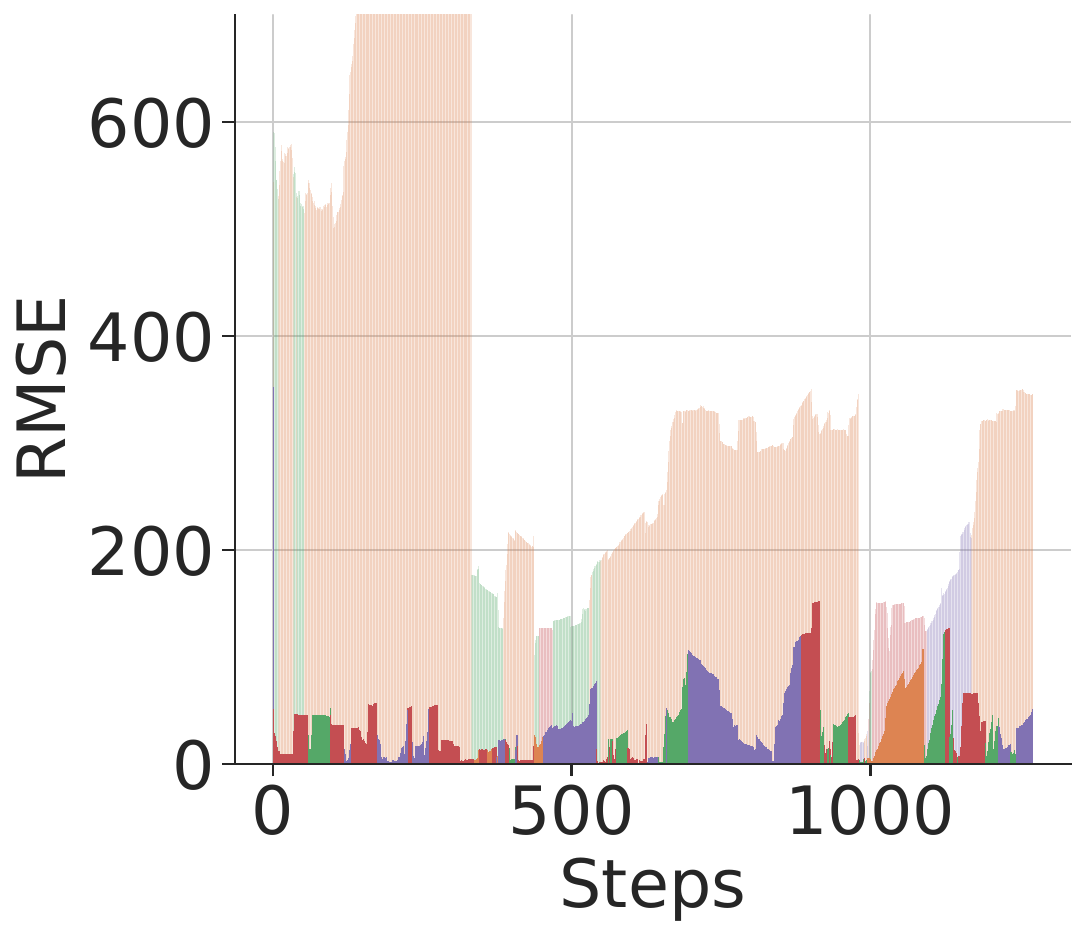}
    \includegraphics[width=0.35\linewidth]{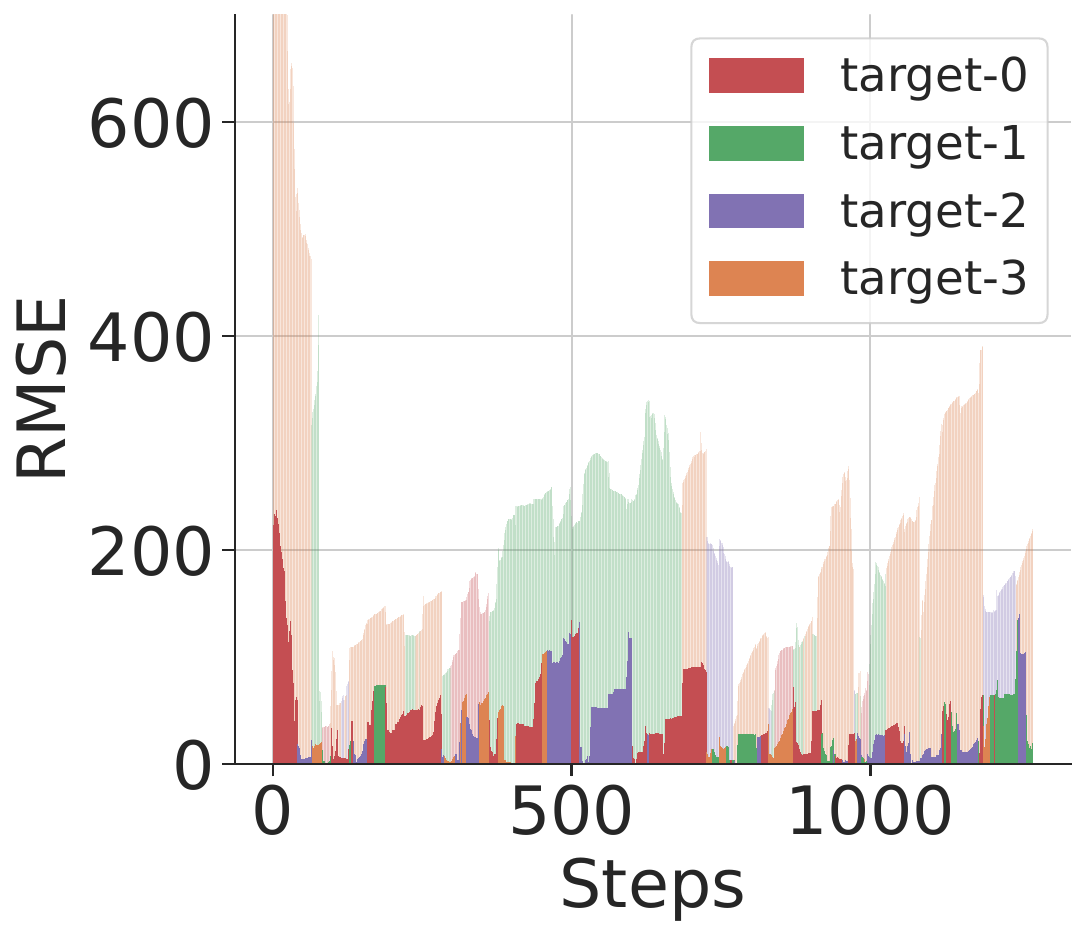}
    \caption{We plot the RMSE of the targets for a scout taking only 2k training steps  to train the NeRF during a planning iteration. The left is of Philadelphia and should be compared against~\cref{fig:RMSE_active} (top right) and the right is of NYC compared against~\cref{fig:RMSE_active} (bottom right).}
    \label{fig:nerf2k}
\end{figure}

Although we quantitatively only see a small difference in PSNR increase from 2k to 4k steps in~\cref{fig:psnr}, we observe a difference in tracking quality. See~\cref{fig:nerf2k} which is the tracking error for the scout that only takes 2,000 training steps per planning step. The upper bounds of the RMSE in both the Philadelphia (left) and NYC (right) are greater than that of the comparative experiments shown in~\cref{fig:RMSE_active} (right). For example in the Philadelphia map, this less trained scout takes longer to find the orange agent and furthermore does a worse job at tracking. We surmise that the better the NeRF representation is, the better its understanding of the occlusions of the scene are, which results is allowing mutual information to extract out poses that view harder to spot blind spots.

\section{Conclusion}
\label{s:conclusion}

In this work we study the pursuit-evasion game in which a scout (quadrotor) must track multiple active ground targets in a large urban scene. We showed that even if we do not have a map, we can train a neural radiance field (NeRF) representation of the scene online and still perform admirably against baselines that use the ground truth map. We saw that the NeRF provides a sufficient representation of past observations and that building an ensemble of NeRFs gives us way to calculate probabilistic information. Furthermore, in order to track targets we use a Bayes filter to represent target locations and we demonstrated how the NeRFs' voxel grid can be used to incorporate this filter. The efficacy of our method provides support that mutual information can methodically combine exploration and tracking objectives. 

Our NeRF is trained with RGB and depth images collected from our OpenGL based simulator that renders Open Street Maps data. We found that reconstruction quality of the scene is important to tracking tasks and further improvements to building the NeRF for larger scenes should improve performance~\cite{xiangli2023bungeenerf, tancik2022blocknerf}. Similarly, in the future we we would like to relax the assumption of having ground truth depth by utilizing monocular depth estimation models such as Marigold~\cite{ke2023repurposing} or DINOv2~\cite{oquab2024dinov2}. 

\section{Acknowledgements}
We would like to thank Bethany Allik, Nathan Schomer, and Franklin Shedleski from ARL for the insightful conversations on this work.

\begin{footnotesize}
\bibliographystyle{ieeetr}
\bibliography{references}
\end{footnotesize}


\end{document}